%% file: technical_report.tex
\documentclass[11pt]{llncs}
\usepackage{t1enc}
\usepackage{pgfplots}
\usepackage{tikz}
\usepackage{fullpage}     
\usepackage{times}

\usepackage{booktabs} 

\usepackage[utf8]{inputenc}
\usepackage{pgfplots}
\usepackage{tikz}
\input{makros.tex}

\usepackage{amsmath}
\usepackage{amssymb}
\usepackage{algorithmic}
\usepackage{algorithm}
\usepackage{multicol}
\usepackage{wrapfig}
\usepackage{graphicx}
\usepackage{numprint}
\npdecimalsign{.} 

\newcommand{\etal}{et~al.}

\newcommand{\quasikernel}{quasi kernel}

\newcommand{\overbar}[1]{\mkern 1.5mu\overline{\mkern-1.5mu#1\mkern-1.5mu}\mkern 1.5mu}
\usepackage{color}
\newif\ifAppendix
\Appendixtrue

\newif\ifFull
\Fulltrue

\usepackage{csvsimple}
\usepackage{soul}

\usepackage[labelformat=simple]{subcaption}
\usepackage{afterpage}

\newcommand{\AlgName}[1]{\ensuremath{\text{{\sf #1}}}}
\newcommand{\lineartime}{\AlgName{LinearTime}}
\newcommand{\nearlinear}{\AlgName{NearLinear}}
\newcommand{\fastker}{\AlgName{FastKer}}
\newcommand{\parfastker}{\AlgName{ParFastKer}}
\newcommand{\redumis}{\AlgName{ReduMIS}}
\newcommand{\vcsolver}{\AlgName{VCSolver}}

\soulregister\cite7
\soulregister\ref7
\soulregister\quasikernel7
\newcommand{\changed}[1]{#1}

\begin{document}
\title{Scalable
  Kernelization for Maximum Independent Sets}
\author{Demian Hespe\inst{1}, Christian Schulz\inst{1,2}, Darren Strash\inst{3}}
\institute{Institute for Theoretical Informatics, Karlsruhe Institute of Technology, Karlsruhe, Germany, \\ \email{\{hespe, christian.schulz\}@kit.edu} \and
University of Vienna, Vienna, Austria, \\\email{christian.schulz@univie.ac.at} \and
Department of Computer Science, Colgate University, Hamilton, NY, USA, \\ \email{dstrash@cs.colgate.edu}}


\maketitle

\begin{abstract}
The most efficient algorithms for finding maximum independent sets in both theory and practice use reduction rules to obtain a much smaller problem instance called a \emph{kernel}. The kernel can then be solved quickly using exact or heuristic algorithms---or by repeatedly kernelizing recursively in the branch-and-reduce paradigm. It is of critical importance for these algorithms that kernelization is fast and returns a small kernel. Current algorithms are either slow but produce a small kernel, or fast and give a large kernel.  We attempt to accomplish both of these goals simultaneously, by giving an efficient parallel kernelization algorithm based on graph partitioning and parallel bipartite maximum matching.\\
We combine our parallelization techniques with two techniques to accelerate kernelization further: dependency checking that prunes reductions that cannot be applied, and reduction tracking that allows us to stop kernelization when reductions become less fruitful.  Our algorithm produces kernels that are orders of magnitude smaller than the fastest kernelization methods, while having a similar execution time. Furthermore, our algorithm is able to compute kernels with size comparable to the smallest known kernels, but up to two orders of magnitude faster than previously possible.
Finally, we show that our kernelization algorithm can be used to accelerate existing state-of-the-art heuristic algorithms, allowing us to find larger independent sets faster on large real-world networks and synthetic instances.
\end{abstract}

%
%




\section{Introduction}
The maximum independent set problem is a classic NP-hard problem~\cite{DBLP:books/fm/GareyJ79} with applications spanning many fields, such as \ifFull classification theory, information retrieval, \fi computer vision~\cite{feo1994greedy}, computer graphics~\cite{sander-mesh-2008}, map labeling~\cite{gemsa2014dynamiclabel} and routing in road networks~\cite{kieritz-contraction-2010}. Given a graph $G=(V,E)$, our goal is to compute a maximum cardinality set of vertices $\mathcal{I}\subseteq V$ such that no vertices in $\mathcal{I}$ are adjacent to one another. Such a set is called a \emph{maximum independent set} (MIS). 
\ifFull %
As a concrete application, independent sets are essential in labeling strategies for maps~\cite{gemsa2014dynamiclabel}, where the objective is to maximize the number of visible non-overlapping labels on a map. 
This problem can solved by constructing the \emph{label
conflict graph}, in which any two conflicting/overlapping labels are connected by
an edge, and then computing a maximum independent set in this graph.
\fi %

One of the most powerful techniques for solving the MIS problem
in practice is \emph{kernelization}---reducing the input to its
most difficult part, the \emph{kernel}. A kernel (for the MIS problem)
of a graph $G$ is a graph
$r(G)$ of smaller or equal size, obtained by applying a specified polynomial time algorithm to $G$ that reduces its
size while preserving the information required to find an MIS in $G$. The algorithm is often composed of
a set of algorithms (so called \emph{reduction} rules), which are applied exhaustively. After
finding a MIS in $r(G)$ we ``undo'' the kernelization to
find an MIS of $G$. Fixed-parameter tractable algorithms for
the MIS problem are exponential in the size of the kernel, and therefore
the MIS problem is considered ``hard'' for a particular instance
when its kernel size is large~\cite{strash2016power}. Thus,
it is often desirable to apply \emph{many} different reduction rules to reduce
the input size as much as possible when solving the problem exactly.

In practice, kernelization is used as a preprocessing step to other
algorithms~\cite{butenko-correcting-codes-2009,conte2017fast,dahlum2016accelerating,sansegundo2016a,strash2016power,verma2015solving}, where speeding up
kernelization directly speeds up the algorithm. However, kernelization may also be
applied repeatedly as part of an algorithm~\cite{akiba-tcs-2016,chang2017computing,lamm2017finding}. 
In either case, the smallest kernels (or seemingly equivalently, the most varied reductions) give the best chance at finding solutions. For instance, the reductions used by Akiba and Iwata~\cite{akiba-tcs-2016} are the \emph{only} ones known to compute an exact MIS on certain large-scale graphs, and \ifFull these reductions\fi{} are further successful in computing exact solutions in an evolutionary approach~\cite{lamm2017finding}. However it is not always beneficial to compute the smallest kernel possible. Fast and simple reductions can compute kernels that are ``small enough'' for local search to quickly find high-quality, and even exact, solutions much faster than the reductions used to find the smallest kernels~\cite{chang2017computing,dahlum2016accelerating}. Fast and simple reductions can even be used to solve many large-scale instances exactly~\cite{strash2016power} just as quickly as the algorithm by Akiba and Iwata~\cite{akiba-tcs-2016}. 

Thus, for kernelization, there is a trade-off between kernel size and kernelization time. The smallest kernels are necessary to solve the most number of instances to optimality, but the fastest reductions have just enough power to solve most instances quickly. %
\ifFull%
However, when run on the largest instances, the large kernels given by simple rules may make it prohibitive to solve these instances exactly, or even near-optimally with heuristic methods. 
\fi%
Thus, to be effective for a majority of applications, kernelization routines should compute a kernel that is as \emph{small} as possible as \emph{quickly} as possible. 



\subsection{Our Results}
To this end, we develop an efficient shared-memory parallel kernelization algorithm based on graph partitioning and parallel bipartite maximum matching.
We combine our parallelization with \emph{dependency checking}---a strategy for pruning
inapplicable reductions---as well as \emph{reduction tracking} that allows us to stop 
kernelization when reductions become less fruitful.
These pruning techniques achieve large additional speedups
over the kernelization of Akiba and Iwata~\cite{akiba-tcs-2016}, which computes similarly sized kernels.
Our experimental evaluation shows that on average our algorithm finds kernels that are seven times smaller
than the algorithms of Chang~\etal~\cite{chang2017computing}, while having similar a running time. At the same time our algorithms are 41 times faster on average than other algorithms that are able to find kernels of~similar~size. In
further experiments we apply our kernelization algorithm to state-of-the-art
heuristic maximum independent set algorithms and find that our kernels can be
used to find larger independent sets faster in large real-world networks and synthetic instances.

\section{Related Work}
\label{sec:relatedwork}
The \emph{maximum clique} and \emph{minimum vertex cover} problems are
equivalent to the maximum independent set problem: a maximum clique in the
complement graph $\overbar{G}$ is a maximum independent set in $G$, and a
minimum vertex cover $C$ in $G$ is the complement of a maximum independent set $V\setminus C$ in $G$. Thus, an algorithm that solves one of these problems can be used to solve the others. Many branch-and-bound algorithms have been developed for the maximum clique problem~\cite{segundo-recoloring,segundo-bitboard-2011,tomita-recoloring}, which use vertex reordering and pruning techniques based on approximate graph coloring~\cite{tomita-recoloring} or MaxSAT~\cite{li-maxsat-2013}, and can be further sped up by applying local search to obtain an initial solution of high quality~\cite{batsyn-mcs-ils-2014}.

A common theme among algorithms for these (and other) NP-hard problems is that of \emph{kernelization}---reducing the input to a smaller instance that, when solved optimally, optimally solves the original instance. Rules that are used to reduce the graph while retaining the ability to compute an optimal solution are called \emph{reductions}. Reductions and kernelization have long been used in algorithms for the minimum vertex cover and maximum independent set problems~\cite{abu-khzam-2007,chen-1999,bodlaender-kernelization-2013,tarjan-1977}, for efficient exact algorithms and heuristics alike.

\subsection{Exact Algorithms} Butenko et al.~\cite{butenko-correcting-codes-2009} and Butenko and Trukhanov~\cite{butenko-trukhanov} were able to find exact maximum independent sets in graphs with thousands of vertices by first applying reductions. Further works have introduced reductions to more quickly solve the maximum clique problem~\cite{sansegundo2016a,verma2015solving} and enumerate $k$-plexes~\cite{conte2017fast}. Though these works apply reduction techniques as a preprocessing step, many works apply reductions as a natural step of the algorithm. Reductions were originally used by Tarjan and Trojanowski~\cite{tarjan-1977} to reduce the running time of the brute force $O(n^22^n)$ algorithm to time $O(2^{n/3})$, and reductions are further used to give the fastest known polynomial space algorithm with running time of $O^*(1.1996^n)$ by Xiao and Nagamochi~\cite{xiao2017exact}. These algorithms apply reductions during recursion, only branching when the graph can no longer be reduced~\cite{fomin-2010}---known as the \emph{branch-and-reduce} method.

Akiba and Iwata~\cite{akiba-tcs-2016} were the first to show the effectiveness of the branch-and-reduce method for solving the minimum vertex cover problem in practice for large sparse real-world graphs. Using a large collection of reductions, they solve graphs with millions of vertices within seconds. In contrast, the vast majority of instances can not be solved by the MCS clique solver~\cite{tomita-recoloring} within a 24-hour time limit~\cite{akiba-tcs-2016}. However, as later shown by Strash~\cite{strash2016power}, many of these same instances can be solved just as quickly by first kernelizing with two simple standard reductions \ifFull(namely, isolated vertex removal and vertex folding reductions) \fi and then running MCS.

\subsection{Heuristic Algorithms}
Kernelization and reductions play an important role in heuristic algorithms too.
Lamm~\etal~\cite{lamm2017finding} showed that including reductions in a branch-and-reduce inspired evolutionary algorithm enables finding exact solutions much faster than provably exact algorithms. \ifFull Dahlum~\etal~\cite{dahlum2016accelerating} further showed how to effectively combine reductions with local search. Dahlum~\etal~find \else Dahlum~\etal~\cite{dahlum2016accelerating} find \fi that standard kernelization techniques are too slow to be effective for local search and show that applying simple reductions in an online fashion improves the speed of local search. Chang~\etal~\cite{chang2017computing} improved on this result, by implementing reduction rules to reduce the lead time for kernelization for local search. 
They introduce two kernelization techniques: a reduction rule to collapse maximal degree-two paths in a single shot, resulting in a fast linear-time kernelization algorithm (\lineartime{}), and a near linear-time algorithm (\nearlinear{}) that uses triangle counting to detect when the domination reduction can be applied. \nearlinear{} has running time $\Oh{\Delta m}$ where $\Delta$ is the maximum degree of the graph.
They further introduce ``reducing--peeling'' to find a large initial solution for local search. This technique can be viewed as computing one path through the search space of a branch-and-reduce algorithm: they repeatedly exclude high-degree vertices and kernelize the graph until it is empty, then take the independent set found as an initial solution for local search.
\ifFull
Their \nearlinear{} algorithm is able to find kernels small enough and fast enough to be effectively used with local search\footnote{Although their implementation of \nearlinear{} has $O(\sqrt{n}m)$ time, as it includes the linear programming reduction.}; however, their kernels are much larger than those of Akiba and Iwata, who use many more advanced reduction rules. Hence, their technique may not be effective for solving large instances exactly.
\fi

\section{Preliminaries}
\label{sec:prelim}
\subsubsection*{Basic Concepts.}
Let  $G=(V,E)$ be an undirected graph on $n = |V|$ nodes and $m = |E|$ edges. We assume that $V=\{0,\ldots, n-1\}$, and to eliminate ambiguity, we at times denote by $V[G]$ and $E[G]$ the sets $V$ and $E$, respectively, for a particular graph $G$.  Throughout this paper, we assume that $G$ is \emph{simple}: it has no multi-edges or self loops.
The set $N(v) = \set{u\mid\set{v,u}\in E}$ denotes the \emph{open} neighborhood (also simply called the \emph{neighborhood}) of $v$.
We further define the open neighborhood of a set of nodes $U \subseteq V$ to be $N(U) = \cup_{v\in U} N(v)$.
We similarly define the \emph{closed} neighborhood as $N[v] = N(v) \cup \{v\}$
and $N[U] = N(U) \cup U$. We sometimes use $N_G$ to denote the neighborhood in a
particular graph $G$.
A graph $H=(V_H, E_H)$ is said to be a \emph{subgraph} of $G=(V, E)$ if $V_H \subseteq V$ and $E_H \subseteq E$. We call $H$ an \emph{induced} subgraph when $E_H = \set{\set{u,v} \in E\mid u,v\in V_H}$.
For a set of nodes $U\subseteq V$, $G[U]$ denotes the subgraph induced~by~$U$.
A set $\mathcal{I} \subseteq V$ of vertices, is said to be an \emph{independent set} if
all nodes in $\mathcal{I}$ are pairwise nonadjacent; that is, $E[G[\mathcal{I}]] = \emptyset$. 
The \emph{maximum independent set problem} is that of finding a maximum cardinality independent set which is called a \emph{maximum independent set} (MIS).

The \emph{graph partitioning problem} is to partition $V$
into $k$ blocks $V_1 \cup \dots \cup V_k = V$ with $V_i \cap V_j = \emptyset, \  \forall i \neq
j$ while optimizing a given cost function---typically the number of edges with
end vertices in different blocks. Additionally, a balance constraint is applied, which demands that the blocks have approximately equal size with respect to the number of vertices or, alternatively, the sum of weights associated with the vertices.
\emph{Boundary vertices} are adjacent to vertices in other blocks and \emph{cut edges} cross block boundaries.

\subsection{Reductions}
\label{subsec:reductions}

We now briefly describe the reduction rules that we consider. Each reduction allows us to choose vertices that are in \emph{some} MIS by following simple rules. If an MIS is found on the kernel graph $\mathcal{K}$, then each reduction may be undone, producing an MIS in the~original~graph.

\subsubsection{Reductions of Akiba and Iwata~\cite{akiba-tcs-2016}.} Akiba and Iwata~\cite{akiba-tcs-2016} use a full suite of advanced reduction rules, which can efficiently solve the minimum vertex cover problem for a variety of instances. Here, we briefly describe the reductions we use, but for the maximum independent set problem. \ifFull Note that Akiba and Iwata further use packing~\cite{akiba-tcs-2016}, and alternative~\cite{Xiao201392} reductions. For brevity, we do not describe them here.\fi

\paragraph*{Vertex folding~\cite{chen-1999}:} For a vertex $v$ with degree two whose neighbors $u$ and $w$ are not adjacent, either $v$ is in some MIS, or both $u$ and $w$ are in some MIS. Therefore, we can contract $u$, $v$, and $w$ to a single vertex $v'$ and decide which vertices are in the MIS later.
  If $v'$ is in the computed MIS, then $u$ and $w$ are added to the independent set, otherwise $v$ is added. Thus, a vertex fold contributes a vertex to an independent set.

\paragraph*{Linear programming relaxation~\cite{nemhauser-1975}:}
  A well-known linear programming relaxation for the MIS
  problem with a half-integral solution (i.e., using only values 0, 1/2, and 1)
  can be solved using bipartite matching: maximize $\sum_{v\in V}{x_v}$ such
  that $\forall (u, v) \in E$, $x_u + x_v \leq 1$ and $\forall v \in V$, $x_v
  \geq 0$. Vertices with value 1 must be in the MIS and can thus be removed from
  $G$ along with their neighbors. {Note that there is a version of this reduction~\cite{iwata-2014}
  that computes a solution whose half-integral part is minimal. However,
  preliminary experiments showed that in practice no additional vertices can be removed.}

\paragraph*{Unconfined~\cite{Xiao201392}:} Though there are several definitions of \emph{unconfined} vertex in the literature, we use the simple one from Akiba and Iwata~\cite{akiba-tcs-2016}. A vertex $v$ is \emph{unconfined} when determined by the following simple algorithm. First, initialize $S = \{v\}$. Then find a $u \in N(S)$ such that $|N(u) \cap S| = 1$ and $|N(u) \setminus N[S]|$ is minimized. If there is no such vertex, then $v$ is confined. If $N(u) \setminus N[S] = \emptyset$, then $v$ is unconfined.  If $N(u)\setminus N[S]$ is a single vertex $w$, then add $w$ to $S$ and repeat the algorithm. Otherwise, $v$ is confined. Unconfined vertices can be removed from the graph, since there always exists a MIS that contains no unconfined vertices.

\paragraph*{Diamond:} Although not mentioned in their paper, Akiba and
  Iwata~\cite{akiba-tcs-2016} extend the
  unconfined reduction in their implementation~\cite{iwata2016pc}. Let $S$ be the set constructed
  in the unconfined reduction for a vertex $v$ that is not unconfined. If there
  are nonadjacent vertices $u_1$, $u_2$ in $N(S)$ such that $N(u_1)\setminus
  N(S) = N(u_2)\setminus N(S) = \{v_1, v_2\}$, then we can remove $v$ from the
  graph because there always exists a MIS that does not contain $v$. Note that this implies that
  $\{v_1, v_2\} \subseteq S$.

\paragraph*{Twin~\cite{Xiao201392}:} Let $u$ and $v$ be vertices of degree three
with $N(u) = N(v)$. If $G[N(u)]$ has edges, then add $u$ and $v$ to
$\mathcal{I}$ and remove $u$, $v$, $N(u)$, $N(v)$ from $G$. Otherwise, some
vertices in $N(u)$ may belong to some MIS $\mathcal{I}$. We still remove $u$,
$v$, $N(u)$ and $N(v)$ from $G$, and add a new gadget vertex $w$ to $G$ with
edges to $u$'s two-neighborhood (vertices at a distance 2 from $u$). If $w$ is
in the computed MIS, then none of $u$'s two-neighbors are in $\mathcal{I}$, and therefore $N(u) \subseteq \mathcal{I}$. Otherwise, if $w$ is not in the computed MIS, then some of $u$'s two-neighbors are in $\mathcal{I}$, and therefore $u$ and $v$ are added to $\mathcal{I}$.


\subsubsection{The Reduction of Butenko~\etal~\cite{butenko-2002}.}
We describe one reduction that was not included in the algorithm by Akiba and Iwata~\cite{akiba-tcs-2016}, but was shown by Butenko~\etal~\cite{butenko-2002} to be highly effective on medium-sized graphs derived from error-correcting codes.

\paragraph*{Isolated Vertex Removal~\cite{butenko-2002}:} If a vertex $v$ forms a single clique $C$ with all its neighbors, then $v$ is called \emph{isolated} (\emph{simplicial} is also used in the literature) and is always contained in some MIS.
  To see this, at most one vertex from $C$ may be in any MIS. Either it is $v$
  or, if a neighbor of $v$ is in an MIS, then we select $v$ instead. Note that
  this reduction rule is completely contained in the
  unconfined reduction rule as every neighbor $u \in N(v)$ of a simplicial vertex $v$
  is unconfined, leaving only $v$ without any neighbors in the graph. As it can be implemented more
  efficiently than the unconfined reduction, we apply the reduction by isolated
  vertex removal before removing unconfined vertices.

\subsubsection{The Linear Time Algorithm by Chang~\etal~\cite{chang2017computing}.}
Chang \etal~\cite{chang2017computing} present a kernelization algorithm \lineartime{} that
runs in time $\mathcal{O}(m)$. It removes vertices of degree zero and one and uses a
reduction rule using maximal paths of degree two. They split the rule into five
cases depending on the length of the maximal degree two path and the endpoints
of the path. The full description can be found in their paper~\cite{chang2017computing}.
The degree two path rule is a specialization of the vertex folding rule explained above, and
does not cover the case of a vertex with two neighbors of degree higher than
two. However, in contrast to the vertex folding rule, it has linear time
complexity.
This algorithm often removes a large fraction of a graphs vertices in
very little time; however, it still leaves the possibility to apply more powerful, but
time consuming reduction rules. We therefore run \lineartime{} as a preprocessing step of
our algorithm.

%


\section{Parallel Kernelization}
\label{sec:parallelization}
As current machines usually have more than one processor and kernelization can
run for hours on large instances, parallelization is a promising way to make larger
graphs feasible for maximum independent set algorithms. In this section, we
describe how we parallelize kernelization: we partition the graph
into blocks so that ``local'' reductions can be run on blocks in parallel, and perform parallel maximum bipartite matching for the ``global''
reduction by linear programming.
Our algorithm first applies the reductions parallelized by partitioning exhaustively.
We then apply the reduction by
linear programming. These steps are repeated until no more vertices can be
removed from the graph. (See pseudocode in Algorithm~\ref{algo:high_level}.)

\begin{algorithm}[h]
\begin{algorithmic}
\STATE{$G \leftarrow$ input graph}
\STATE{$\{V_1,\dots , V_k \} \leftarrow partition(G, k)$}
\WHILE{$G$ changed in last iteration}
\FORALL{blocks $V_i$ in parallel}
\STATE{$G \leftarrow localReductions(G, V_i)$}
\ENDFOR
\STATE{$G \leftarrow parallelLinearProgrammingReduction(G)$}
\ENDWHILE
\end{algorithmic}
\caption{Algorithm Overview}
\label{algo:high_level}
\end{algorithm}

\subsection{Blockwise Reductions}
Many reductions have an element of locality. In particular, we call a reduction \emph{local} if it is applied one vertex at a time, if determining that the reduction can be applied is based on local graph structure (for example, by its neighborhood or by neighbors of neighbors), and the reduction itself modifies only local graph structure.
A challenge in parallelizing local reductions is in how to apply them simultaneously.
Fine-grained parallelism would require locks, since attempting to simultaneously remove or contract (near-)neighboring vertices in the graph results in a race condition:
these (near-)neighbors may \emph{both} be mistakenly added to the independent set or the graph may be modified incorrectly.
However, with locks, local reductions become more expensive---reductions must wait if they overlap other reductions in progress.

To avoid locks altogether, we partition the graph into vertex-disjoint blocks and perform local reductions on each block in parallel (i.e., \emph{blockwise}). Note that the only way for two blockwise reductions to simultaneously (mistakenly) reduce neighbors is if they are incident to a cut edge.
We therefore avoid race conditions by restricting reductions to only read and write to vertices and neighborhoods within
a single block. {As reductions may still be applied,
  we call the resulting graph a \quasikernel{} instead of a kernel.}
By using a high-quality partitioning that minimizes the number of cut edges,
we expect the number of vertices excluded from these local
reductions to be small. To avoid race conditions when removing boundary vertices from
the graph, we leave the adjacency lists of neighboring vertices unchanged {and
only mark vertices as removed from the graph}.
We only change the adjacency list of vertices when performing vertex contractions.

We now explain how to apply each local reduction in our parallel framework. Let $V_i$ be
the block in which we are applying the reduction.
\ifFull
Let a reduction on a vertex $v
\in V_i$ only depend on (and modify) vertices $R(v)\subseteq V_i$.
Then no other vertex $u \in V_j$ for all $i \neq j$ is traversed or modified as the result
of this reduction. Thus, we apply this reduction to $v$ correctly: changes to it and/or adjacency lists
of vertices in $R(v)$ do not affect vertices of other blocks.

\fi
Further, let $B_k$ denote the set of vertices of distance at most $k$ from
some boundary vertex in our partitioning. Note that $B_0$ is the set of boundary vertices and $B_k = N[B_{k-1}]$.
\paragraph*{Vertex Folding:} 
Let $v\in V_i$ be a vertex with neighborhood $\{u,w\}\subset V_i$. Contracting
$v,u,w$ into $v'$ will cause a race condition whenever $u,w\in V_i\cap B_0$, as
their neighbors in some other block $V_j$ must have their adjacency lists
updated to include $v'$. {In these cases, we do not apply the vertex folding reduction. }We handle vertex folding with two cases. First, for $u,w\in V_i\setminus B_0$ (or equivalently, $v\in V_i\setminus B_1$), we apply the reduction normally. Then $N(\set{u,w})\subseteq V_i$ and there is no race condition. Secondly, without loss of generality, if $u\in V_i\cap B_0$ and $w\in V_i\setminus B_0$ then we still apply vertex folding, using $u$ as the new vertex $v'$. Neighborhoods of vertices in \ifFull $N(u) \setminus \{v\}\not\subseteq V_i$\else $N(u) \setminus \{v\}$\fi{} remain unchanged.

\begin{figure}[t]
  \centering
  \includegraphics[scale=1.3]{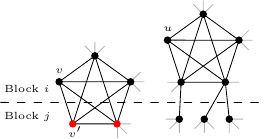}
  \caption{Vertices $v$ and $v'$ are simplicial but lie in different
    blocks, they are not removed from the graph. Vertex $u$ is simplicial and
  is not a boundary vertex, so $N[{u}]$ is removed.}
  \label{fig:parallel_isolated_clique}
\end{figure}

\paragraph*{Isolated Vertex Removal:} Let $v\in V_i\setminus B_0$ be an isolated
vertex. Then we add $v$ to $\mathcal{I}$ and remove $N[v]\subseteq V_i$ from the graph as usual. (See Figure~\ref{fig:parallel_isolated_clique}.)

\paragraph{Twin:} Let $u,v\in V_i$ such that $N(u) = N(v)\subseteq V_i$, and note that $u,v$ will not be boundary vertices as otherwise $N(u)\not\subseteq V_i$. We have two cases:
\begin{description}
  \item[{$G[N(u)]$ has edges}:] Since $u,v$ are not boundary vertices, we add $u,v$ to $\mathcal{I}$ and remove $\{u,v\}\cup N(u)\subseteq V_i$ from the graph.
  \item[{$G[N(u)]$ has no edges}:] We only apply this reduction when $N[N[u]]\subseteq V_i$: we remove $\{u,v\}\cup N(u)$, and create a new vertex $w\in V_i$ with neighborhood $N(w) = N[N[u]]$; otherwise we would modify the adjacency list of a vertex in a different block. 
  \end{description}
  \paragraph*{Unconfined:} Unlike other blockwise reductions, every vertex $v\in V_i$ is eligible for the unconfined reduction, including boundary vertices. If a vertex is unconfined, we mark it as excluded from the independent set and remove it from the graph (by setting a flag if $v$ is a boundary vertex).
 However, the algorithm for finding
    unconfined vertices must be adapted---it does not simply rely on a (two-)neighborhood, but depends on an expanding set of vertices $S$, which should be drawn from $V_i$ in order to avoid a race condition. In particular, a vertex $u \in N(S)$ can only be used if $u\in V_i$ and $S\subseteq V_i$ must hold.
    This way, we ensure all vertices that we classify as unconfined
    are truly unconfined and can be removed from the graph. We might, however,
    falsely classify some vertices as confined.
  \paragraph*{Diamond:} As with the unconfined reduction, we can safely remove even
    boundary vertices from the graph by using the diamond reduction. However, since vertices in $V\setminus V_i$
    cannot be inserted into $S$ during the blockwise unconfined reduction, there might
    be $u_1, u_2$ such that $N(u_1) \setminus N(S) = N(u_2) \setminus N(S) =
    \{v_1, v_2\}$ and $\{v_1, v_2\} \not\subseteq S$ because they are located in
    different blocks, so we have to check that $v_1, v_2 \in S$. If 
    not, they might be removed by another reduction which can lead to race conditions.

\subsection{Parallel Linear Programming}
Unlike the local reductions, the reduction by linear programming is not applied to single vertices and their (near-)neighbors.
It instead relies on a \emph{global view} of 
the graph to find a set of vertices that can be removed at once. Therefore our
parallelization strategy for local reductions cannot be applied to the linear programming reduction. The computationally
expensive part of this reduction is finding a maximum
bipartite matching of the bi-double graph: $B(G) = (L_V \cup R_V, E')$, where
$L_V = \{l_v \mid v \in V\}, 
R_V  = \{r_v \mid v \in V\}, \text { and } 
E'=\{\{l_u,r_v\}\mid\{u,v\}\in E\}$.

Azad et al.~\cite{azadcomputing} give a parallel augmenting path based
algorithm for maximum bipartite matching. Their algorithm requires a maximal matching as input, which we first compute using the maximal matching algorithm by Karp and Sipser~\cite{karp1981maximum}, which was parallelized by Azad et al.~\cite{azad2012multithreaded}.
For better performance when repeatedly applying the reduction, we reuse the parts of the previous matching which are still part of the graph. If the graph changed only slightly since the last application, this is still close to a maximum matching, which results in less work for the maximum matching algorithm. This technique is also used by Akiba and Iwata~\cite{akiba-tcs-2016}.
To obtain the half-integral result of the linear program, we use the set of
vertices reachable by alternating paths starting from matched vertices in $L_V$.
To find these, we start a depth first search from each vertex $v \in L_V$ in
parallel and mark all reached vertices. We then obtain the result by
iterating in parallel over all vertices in the original graph and checking whether their
respective vertices in $L_V$ and $L_R$ are marked.

\section{Pruning Reductions}
\label{sec:pruning}
\subsection{Dependency Checking}
\label{sec:dependency}

To compute a kernel, Akiba and Iwata~\cite{akiba-tcs-2016} apply their
reductions~$r_1, \dots ,r_j$ by iterating over all reductions and trying to
apply the current reduction $r_i$ to all vertices. If $r_i$ reduces at
least one vertex, they restart with reduction~$r_1$. When reduction~$r_j$ 
is executed, but does not reduce any vertex, all reductions have been applied
exhaustively, and a kernel is found. Trying to apply every reduction to all
vertices can be expensive in later stages of the algorithm where 
few reductions succeed. The algorithm may repeatedly
attempt to apply the same reduction to a vertex even though
the graph has not changed sufficiently to allow the reduction to succeed. For example, let
$G'$ be a graph obtained by applying reductions to a graph $G$. If
vertex $v$ is not isolated in $G$ and $N_{G'}[v] = N_G[v]$, then $v$ is still not isolated in $G'$ and can be pruned from further attempts.

\begin{wrapfigure}[15]{r}[0cm]{5cm}
  \centering
  \includegraphics[scale=1.2]{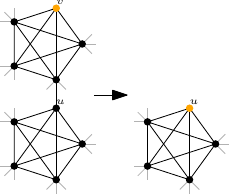}
  \caption{After removing isolated vertex $v$ and $N(v)$, $u$ is isolated. Orange vertices are
    in $D$.}
  \label{fig:dependency_checking}
\end{wrapfigure}
We define a scheme for checking dependencies between reductions, which allows us to
avoid applying isolated vertex removal, vertex folding, and twin reductions when they
will provably not succeed. After unsuccessfully trying to apply one
of these reductions to a vertex $v$, one only has to consider $v$ again for reduction
after its neighborhood has changed. We therefore keep a
set $D$ of \emph{viable} candidate vertices: vertices whose neighborhood has changed
and vertices that have never been considered for reductions. Initially we set $D = V$.
Then for each $v \in D$, we remove $v$ from $D$ and try to apply our reductions to $v$.
If $v$ is removed from the graph (or a new vertex $w$ is inserted),
we set $D = D \cup N(v)$ (or $D = D \cup N[w]$). We repeat until $D$ is empty.
Figure~\ref{fig:dependency_checking} shows an example\ifFull for isolated vertex
removal.\else.\fi

\ifFull
Using this technique we reduce the amount of work for kernelization, especially
in the later stages of the algorithm, where only few reductions are left to
apply. 
Dependency checking can also help finding a kernel faster after finding a
\quasikernel{} using our parallel algorithm: as most parts of the graph are already fully reduced, we expect
dependency checking to quickly prune these parts and focus further kernelization
on the boundaries when running the sequential version of our algorithm on the \quasikernel.
\fi
Note that this strategy does not support 
unconfined and diamond reductions, as they depend on a set $S$ that can
\ifFull grow arbitrarily large, and\fi{}include vertices with large distances from the starting vertex. Thus
a vertex can become unconfined due to a change in the graph outside of its
neighborhood. Neither does it support the linear programming reduction, which operates 
on the entire graph instead of a single vertex.
However, when performing these reductions we continue to add vertices
whose neighborhoods have changed to $D$, saving
effort when next attempting isolated vertex removal, vertex folding, and twin reductions.

We briefly mention that targeted forms of dependency checking have been used before. Previous works, including Akiba and Iwata~\cite{akiba-tcs-2016} and Chang~\etal~\cite{chang2017computing} perform so-called ``iterated'' reductions, which allow for the repeated application of successful reductions. These include, for example, iteratively removing degree-one and -zero vertices \ifFull(including any newly introduced degree-one and -zero vertices)\fi{} until none remain, and applying the domination reduction when triangle counts change~\cite{chang2017computing}. Unlike these previous works, our focus is on eliminating reductions that cannot be applied, as is not targeted at any particular reduction, but a collection of reductions. Strash~\cite{strash2016power} implements similar dependency checking for isolated vertex and vertex folding reductions, though it is not mentioned in his paper. We are the first to introduce such a strategy that can be used with \emph{any} collection of reductions.

\subsection{Reduction Tracking: Counteracting Diminishing Returns}
\label{sec:diminishing}

It is not \emph{always} ideal to apply reductions exhaustively---for example if only few reductions will succeed and they are costly.
We note that, during later stages of our algorithm, local reductions may lead to very few graph changes, 
while the linear programming reduction often significantly reduces the graph size. Therefore, it may be better to stop local reductions early before applying the
linear programming reduction, as any remaining local reductions can still be applied afterwards.
Furthermore, in our parallel algorithm, applying the local
reductions exhaustively can take significantly longer for some blocks than for
others. That is, the total graph size is not significantly reduced once the first threads finish their blocks.

We therefore implement reduction tracking to detect and stop local reductions when they are not quickly reducing the graph.
Once the first thread finishes applying local reductions, we assign it to sample the current graph size at fixed time intervals. 
We then stop local reductions on all threads when the change in graph size becomes
small relative to the rate of vertex removals and switch to the linear programming reduction.
We continue local reductions afterwards.
For the sequential case, we start sampling the current size immediately when starting the local
reductions. In our implementation, sampling is performed by an
  additional thread; however, it does not introduce significant overhead and can be
  done in the same thread.


\section{Experimental Evaluation}
\label{sec:experiments}

\paragraph*{Methodology.} We implemented our algorithm using C++ and compiled all
code using gcc~5.4.0 with full optimizations turned on (\texttt{-O3}~flag). For shared
memory parallelization we use OpenMP~4.0. Our implementation includes the
parallel application of reduction rules, the dependency checking scheme and the
reduction tracking technique. Our source code is available on
github\footnote{\url{https://github.com/Hespian/ParFastKer}} and a sequential version
of our algorithm has been integrated into the KaMIS software for finding high
quality independent sets\footnote{\url{http://algo2.iti.kit.edu/kamis/}}.
For graph partitioning we use ParHIP~\cite{meyerhenke2017parallel}, the parallel version of the KaHIP graph partitioner~\cite{kabapeE}.
We compare against several existing sequential kernelization techniques. 
For fast reduction strategies, we compare against the kernelization routines \lineartime{} and \nearlinear{} recently introduced by
Chang~\etal~\cite{chang2017computing}. We use the authors' original implementation, written
in C. For extensive reduction strategies, we use the full reduction suite of Akiba and
Iwata's \vcsolver{}~\cite{akiba-tcs-2016}\footnote{\url{https://github.com/wata-orz/vertex\_cover}}.
We modified their code to stop execution after
kernelization and output the kernel size.
For all instances, we
perform three independent runs of each algorithm.
Their code was compiled and run sequentially with Java~1.8.0\_102.
All results are averages over three runs on a machine
with 512 GB RAM and two Intel Xeon E5-2683 v4 processors with 16 cores
running at 2.1 GHz each.

  \paragraph*{Data Structure Details.} We represent our graph using adjacency
  lists: For every vertex, we store an array of its neighbors ids. When a vertex
  is removed from the graph, it is not removed from the adjacency lists of its
  neighbors, but instead is marked as removed in an array accessible by all
  threads. Note that we cannot store all edges of the graph consecutively as the
  vertex folding and twin reductions can increase the number of neighbors of a
  vertex.  To efficiently check
  the degree of a vertex or it's number of incident cut edges, we store these
  values using atomic integers.

  For each block, we additionally store the following data structures that are
  only used by the thread that is handling the respective block.
In order to efficiently iterate over vertices that have not been
  removed from the graph, we keep a consecutive array of vertex ids from the
  respective block. For
  constant time removal of vertices from this array, we store
  additional pointers from the vertex id to the position in the array. This data
structure is also used in our dependency checking technique to store the
vertices that have to be considered for reduction.



\paragraph*{Algorithm Configuration.} We run our
algorithm with all reduction rules explained in this paper but restrict the
isolated vertex removal reduction to cliques of size 3 or less. We use the
\emph{ultrafast} configuration of the parallel partitioner and default values
for all other parameters. When running our algorithm in parallel on $p$ threads,
we partition the graph into $p$ blocks. We stop applying local reduction rules
when the reduction in graph size per time during the last time interval is less than $5\%$ of
the average size reduction per time since starting to apply local reductions
(i.e., since the last application of the linear programming reduction). An
experimental evaluation of this technique can be found in \changed{Section~\ref{sec:reductionStopping}}.
As the \lineartime{} algorithm by Chang~\etal~\cite{chang2017computing} has very
low running times and reduces the initial graph size, we run it as a preprocessing step
using the original implementation. We then partition the resulting kernel and process
it with our parallel kernelization algorithm.
Throughout this section, we will refer to sequential runs of our algorithm as
\fastker{} and to parallel runs (32 threads, unless otherwise stated) as
\parfastker{}. All repetitions of \parfastker{} use the same partitioning of the input.

\paragraph*{Instances.}
We perform experiments on large web~\cite{BCSU3} and road
networks~\cite{bader2014benchmarking,demetrescu2009shortest}, random
(hyper)-geometric graphs~\cite{kappa,von2016generating} and Delaunay
triangulations~\cite{bader2014benchmarking,meyerhenke2017parallel}. 
Basic instance properties can be found in Table~\ref{tab:graphproperties}.
These instances are all large ($> 10$M
  vertices) and kernelization takes a considerable amount of time on them. As
  our methods introduce some overhead compared to other kernelization
  algorithms, we focus our attention on speeding up kernelization for these hard instances.
\begin{table}[H]
\center
\begin{tabular}{ll|rr|r}
name & type & \# vertices & \# edges & from \\ 
\hline 
\hline 
uk-2002 & web & \numprint{18.5}M & \numprint{261.8}M & \cite{bader2014benchmarking} \\
arabic-2005  & web      & \numprint{22.7}M  & \numprint{553.9}M  & \cite{BCSU3} \\
gsh-2015-tpd & web      & \numprint{30.8}M  & \numprint{489.7}M  & \cite{BCSU3} \\
uk-2005      & web      & \numprint{39.5}M  & \numprint{783.0}M  & \cite{BCSU3} \\
it-2004      & web      & \numprint{41.3}M  & \numprint{1027.5}M & \cite{BCSU3} \\
sk-2005      & web      & \numprint{50.6}M  & \numprint{1810.1}M & \cite{BCSU3} \\
uk-2007-05   & web      & \numprint{105.9}M & \numprint{3301.9}M & \cite{BCSU3} \\
webbase-2001 & web      & \numprint{118.1}M & \numprint{854.8}M  & \cite{BCSU3} \\
        \hline
asia.osm     & road     & \numprint{12.0}M  & \numprint{12.7}M   & \cite{bader2014benchmarking} \\
road\_usa    & road     & \numprint{23.9}M  & \numprint{28.9}M   & \cite{demetrescu2009shortest} \\
europe.osm   & road     & \numprint{50.9}M  & \numprint{54.1}M   & \cite{bader2014benchmarking} \\
        \hline
rgg26        & rgg & \numprint{67.1}M  & \numprint{574.6}M  & \cite{bader2014benchmarking} \\
rhg          & rhg & \numprint{100.0}M & \numprint{1999.5}M & \cite{von2016generating} \\
        \hline
del24        & delaunay & \numprint{16.8}M  & \numprint{50.3}M   & \cite{bader2014benchmarking} \\
del26        & delaunay & \numprint{67.1}M  & \numprint{201.3}M  & \cite{meyerhenke2017parallel} \\
\end{tabular}

\vspace*{.25cm}
\caption[Graphs used]{Basic properties of the graphs used in our evaluation.}
\label{tab:graphproperties}
\end{table}

\subsection{Comparison with State-of-the-Art}
\label{sec:comparison}

We now compare our implementation to the implementations of \vcsolver{} by Akiba and
Iwata~\cite{akiba-tcs-2016} and the \lineartime{} and \nearlinear{} algorithm by
Chang~\etal~\cite{chang2017computing}. 
Table~\ref{tab:summary} and Figure~\ref{fig:summaryPlot} give an overview. Figure~\ref{fig:summaryPlot} normalizes running time and kernel size on each instance by the result of
\vcsolver{}.

First note that \lineartime{}'s running time is almost negligible compared to
that of \vcsolver{}, almost never surpassing $1\%$ of \vcsolver{}'s
time. \lineartime{} also decreases the graph size significantly for most graphs
(except for the Delaunay triangulations, where \lineartime{} is not able to
reduce the graph size at all), however, the \lineartime{} kernel is still orders
of magnitude larger than \vcsolver{}'s kernel. Due to fast running time and
graph size reduction, we use \lineartime~as a preprocessing
step to our algorithm.

\begin{figure*}[!h]
  \centering
\footnotesize
\begin{tabular}{lr|rr|rr|rr|rr|rrr}%
 \multicolumn{2}{c|}{Graph} & \multicolumn{2}{c|}{\lineartime} & \multicolumn{2}{c|}{\nearlinear{}} & \multicolumn{2}{c|}{\vcsolver{}} & \multicolumn{2}{c|}{\fastker{}} & \multicolumn{3}{c}{\parfastker} \\
 name & $n$ & $|\mathcal{K}|$ & time & $|\mathcal{K}|$ &  time & $|\mathcal{K}|$ & time & $|\mathcal{K}|$ & time & $|\mathcal{K}|$ & time & su \\
  \hline
    \begin{filecontents*}{include/allGraphsComparison.csv}
graph,graphSize,LinearTimeTime,LinearTimeSize,NearLinearTime,NearLinearSize,AkibaTime,AkibaSize,OurSequentialTime,OurSequentialSize,OurParallelTime,OurParallelSize,ParallelSpeedupAkiba
uk-2002,\numprint{19}M,\numprint{1.5},\numprint{11.7}M,\numprint{28.0},\numprint{4.0}M,\numprint{336.9},\numprint{0.2}M,\numprint{60.1},\textbf{\numprint{0.3}M},\numprint{11.8},\textbf{\numprint{0.3}M},\numprint{28.4}
arabic-2005,\numprint{23}M,\numprint{2.6},\numprint{15.6}M,\numprint{246.1},\numprint{6.7}M,\numprint{1033.2},\numprint{0.6}M,\numprint{148.0},\textbf{\numprint{0.6}M},\numprint{25.7},\textbf{\numprint{0.6}M},\numprint{40.2}
gsh-2015-tpd,\numprint{31}M,\numprint{11.6},\numprint{2.0}M,\numprint{97.4},\numprint{1.2}M,\numprint{372.3},\numprint{0.4}M,\numprint{66.4},\textbf{\numprint{0.4}M},\numprint{32.0},\textbf{\numprint{0.5}M},\numprint{11.7}
uk-2005,\numprint{39}M,\numprint{2.5},\numprint{28.2}M,\numprint{60.5},\numprint{5.9}M,\numprint{541.4},\numprint{0.8}M,\numprint{131.9},\textbf{\numprint{0.9}M},\numprint{53.3},\textbf{\numprint{0.9}M},\numprint{10.1}
it-2004,\numprint{41}M,\numprint{3.3},\numprint{27.1}M,\numprint{1544.6},\numprint{11.3}M,\numprint{6749.0},\numprint{1.6}M,\numprint{499.7},\textbf{\numprint{1.7}M},\numprint{151.8},\textbf{\numprint{1.7}M},\numprint{44.4}
sk-2005,\numprint{51}M,*,\textbf{*},*,\textbf{*},\numprint{10010.5},\numprint{3.2}M,\numprint{2349.8},\textbf{\numprint{3.3}M},\numprint{178.3},\numprint{3.5}M,\numprint{56.1}
uk-2007-05,\numprint{106}M,*,\textbf{*},*,\textbf{*},\numprint{18829.4},\numprint{3.5}M,\numprint{2073.4},\textbf{\numprint{3.6}M},\numprint{372.4},\textbf{\numprint{3.7}M},\numprint{50.6}
webbase-2001,\numprint{118}M,\numprint{13.0},\numprint{51.7}M,\numprint{121.1},\numprint{17.3}M,\numprint{4207.8},\numprint{0.7}M,\numprint{290.8},\textbf{\numprint{0.8}M},\numprint{54.9},\textbf{\numprint{0.9}M},\numprint{76.6}
asia.osm,\numprint{12}M,\numprint{0.8},\numprint{626.7}K,\numprint{1.4},\numprint{594.4}K,\numprint{204.7},\numprint{15.2}K,\numprint{1.6},\textbf{\numprint{34.9}K},\numprint{1.2},\textbf{\numprint{34.9}K},\numprint{169.8}
road\_usa,\numprint{24}M,\numprint{2.5},\numprint{2.5}M,\numprint{4.1},\numprint{2.4}M,\numprint{310.0},\numprint{0.2}M,\numprint{8.0},\textbf{\numprint{0.2}M},\numprint{4.1},\textbf{\numprint{0.2}M},\numprint{76.0}
europe.osm,\numprint{51}M,\numprint{4.1},\numprint{1500.0}K,\numprint{6.1},\numprint{1329.9}K,\numprint{302.4},\numprint{8.4}K,\numprint{5.8},\textbf{\numprint{14.1}K},\numprint{4.9},\textbf{\numprint{14.2}K},\numprint{61.3}
rgg26,\numprint{67}M,\numprint{1.0},\numprint{67.1}M,\numprint{172.6},\numprint{51.3}M,\numprint{9887.7},\numprint{49.6}M,\numprint{13572.6},\textbf{\numprint{49.8}M},\numprint{150.3},\textbf{\numprint{49.8}M},\numprint{65.8}
rhg,\numprint{100}M,*,\textbf{*},*,\textbf{*},\numprint{124.0},\numprint{0},\numprint{164.5},\textbf{\numprint{0}},\numprint{64.6},\textbf{\numprint{16}},\numprint{1.9}
del24,\numprint{17}M,\numprint{0.2},\numprint{16.8}M,\numprint{12.7},\numprint{15.6}M,\numprint{4789.5},\numprint{12.4}M,\numprint{142.0},\numprint{12.9}M,\numprint{51.5},\numprint{12.9}M,\numprint{93.1}
del26,\numprint{67}M,\numprint{0.7},\numprint{67.1}M,\numprint{53.3},\numprint{62.5}M,\numprint{20728.7},\numprint{49.9}M,\numprint{718.9},\numprint{51.7}M,\numprint{179.0},\numprint{51.7}M,\numprint{115.8}
\end{filecontents*}
\csvreader[head to column names, late after line=\\]{include/allGraphsComparison.csv}{}
    {\graph & \graphSize &\LinearTimeSize & \LinearTimeTime & \NearLinearSize & \NearLinearTime & \AkibaSize & \AkibaTime & \OurSequentialSize & \OurSequentialTime & \OurParallelSize & \OurParallelTime & \ParallelSpeedupAkiba}
    \end{tabular}

\hspace*{-1cm}
    \caption{Running times and kernel sizes ($|\mathcal{K}|$) for all algorithms. The column ``su'' is the speedup of \parfastker{} over \vcsolver{}. Instances marked with a star (*) cannot be processed by the
      \nearlinear{} and \lineartime{}
      implementations due to the 32-bit implementation. All times are in
      seconds. Quasi kernel sizes that differ from \vcsolver{}'s kernel size by at most $0.5 \%$ of the graph
      size are emphasized in \textbf{bold}.}
    \label{tab:summary}
\end{figure*}

The \nearlinear{} algorithm by Chang~\etal~\cite{chang2017computing} uses fewer
reduction rules than our algorithm, so it finds larger kernels, often
orders of magnitude larger than the kernels by \vcsolver{} and our algorithms.
The largest relative difference to the smallest kernel size of \nearlinear{} is
the \numprint{1329923}-vertex kernel for \texttt{europe.osm}. This is 159 times
larger than the smallest kernel and 94 times larger than the \quasikernel{} found by
\parfastker{}.
For the Delaunay triangulations and the random geometric graph, the relative
kernel size difference is comparatively low. This is
because the kernel for these graphs is still very large compared to the input
size, but we find \quasikernel{}s much closer to the size found by \vcsolver{}
than \nearlinear{}. \lineartime{} actually cannot remove any vertices from the
Delaunay instances and only very few from the random geometric instance.
In the geometric mean, \lineartime{}'s kernel is a factor 12 larger than
\parfastker{}'s \quasikernel{} and 
\nearlinear{}'s kernel is a factor 7 larger. 
Due to \nearlinear{}'s fast worst-case running time, it runs
faster than \fastker{} on 8 out of 12 instances and on 2 instances even faster
than \parfastker{}. As \lineartime{} is a preprocessing step of our algorithm, it is
of course always faster.
\afterpage{
\begin{figure*}[!tb]
\centering
\begin{minipage}[t]{.283\textwidth}
\centering
\includegraphics[width=\textwidth]{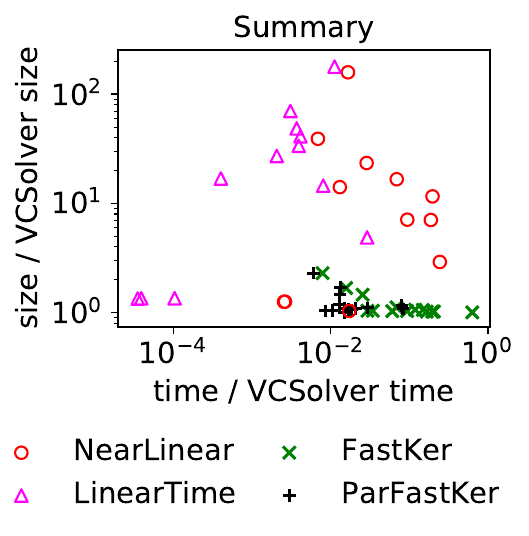}
\caption{Comparison of kernelization algorithms against \vcsolver{}.\protect\footnotemark}
\label{fig:summaryPlot}
\end{minipage}
\hspace{.005\textwidth}
\begin{minipage}[t]{.70\textwidth}
\centering
\includegraphics[width=\textwidth]{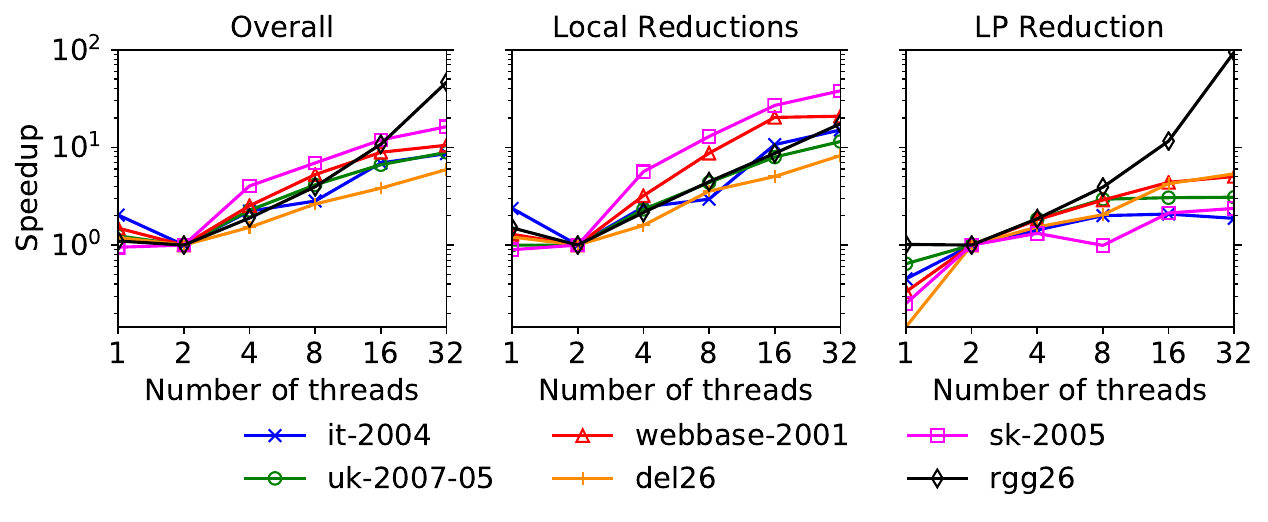}
  \caption{Scaling experiments on the six hardest instances of
    the benchmark set for (left) the overall algorithm,
     (center) blockwise reductions and (right) the reduction by linear programming. Speedups are relative to two threads.}
  \label{fig:scaling}
\end{minipage}
\end{figure*}
\footnotetext{As empty kernels lead to a division by zero for this plot, graphs
  with an empty kernel are not shown here.}
}

As \vcsolver{} implements a larger set of reduction rules, adding the desk and funnel
reductions by Xiao and Nagamochi~\cite{Xiao201392} as well as Akiba and Iwata's own \emph{packing}
reduction rule, it achieves smaller kernel sizes. In the geometric mean, \parfastker{}'s
\quasikernel{}'s are $20\%$ larger than \vcsolver{}'s (excluding \texttt{rhg},
which has an empty kernel).
However, comparing the kernel
sizes to the size of the input network, these differences in size are
negligible. The largest obtained difference relative to the size of the input
network among all graphs we tested, is $2.9\%$ on \texttt{del24} ($0.6\%$ on \texttt{sk-2005} when only
considering the real-world instances). \changed{In addition, \vcsolver{} only applies a scheme similar to our dependency checking for the removal of degree zero and one
vertices, so our algorithm runs faster on all instances except \texttt{rhg} and \texttt{rgg26}. \parfastker{}, however, is faster than \vcsolver{} on these
instance.} On 11 out of 15 instances, \fastker{} is faster by
a factor of over \changed{5} than \vcsolver{} and on \changed{5 instances even by a factor of over 28}. The largest speedup of \fastker{} over \vcsolver{}
is \changed{$129$} on \texttt{asia.osm} and the geometric mean of the speedups is \changed{10}. As
\fastker{} is the sequential version of our algorithm, this is a purely
algorithmic speedup. Using
parallelization, \parfastker{} achieves speedups of 41 over
\vcsolver{} in the geometric mean, combining the algorithmic speedup with
parallel speedup. On all instances, except for \texttt{rhg},
the speedup is over 10 and on 9 instances over 50.

\subsection{Scalability}
\label{sec:scaling}
Figure~\ref{fig:scaling} shows the parallel speedup of our algorithm on the six hardest
instances of our benchmark set (i.e., those with the longest sequential running time).
The left plot shows the total speedup relative to two threads for all parts of our
algorithm combined: \lineartime{} preprocessing, partitioning and parallel
reductions with dependency checking and inexact reduction pruning.
The center and right plots show
the speedups for the reductions parallelized by partitioning and
the reduction by linear programming, respectively. The preprocessing step of our
algorithm, the \lineartime{} algorithm by Chang~\etal, is sequential and thus limits the
possible scalability of our parallelization, however running times are very short.

We observe that, due to the overhead caused by having to find a partition of the
graph, the single threaded execution is on average \changed{1.7} times faster than the 
parallelization using 2 threads. However, our algorithm scales well so that parallelization brings better performance for
higher numbers of threads.
Compared to the two-threaded case, our highest speedup is 46.5
for \texttt{rgg26} on 32 threads. The main reason for this is that
reductions on this graph are so slow that, for low thread counts, our inexact reduction pruning
technique stops local reductions early, switching to a very long lasting
reduction by linear programming. For the other graphs, the speedup on 32 threads
compared to 2 threads is between 6 and 16.3 with 16 being perfect speedup. \changed{The
speedup relative to the single threaded case is
between~3.3~and~13.1~(42.1~for~\texttt{rgg26}).}

Figure~\ref{fig:scaling} shows that 
local reductions parallelized by partitioning are faster single threaded
than on two threads. This is caused by our inexact reduction pruning technique which starts
after the first thread finishes reductions. When the number of threads is low,
reductions might already have become too slow when the first thread finishes,
causing longer times of slow size reduction. For higher thread counts, there is
always a thread that finishes while other threads are still applying reductions
fast and thus less time is wasted by slow reductions. After the drop at two
threads, the speedup for 32 threads compared to 2 threads for these reductions
is between 8~and~37 \changed{(between~4.8~and~32 compared to~1~thread)}.
For some graphs, the reduction by linear
programming, which we parallelized using the parallel maximum bipartite matching
algorithm by Azad~\etal~\cite{azadcomputing}, is a bottleneck of our algorithm
{as it does not scale as well as the rest of the reductions. In many
  cases, about half of the reduction time is spend on this reduction rule alone.}

\subsection{Reduction Tracking: Counteracting Diminishing Returns}
\label{sec:reductionStopping}
     
Our experiments show that stopping long lasting reductions early can lead to
significant speedups \changed{on some graphs} with close to no penalty on the \quasikernel{} size. \changed{The
\quasikernel{} size found with reduction tracking enabled is less
than $0.1\%$ larger than without it on all but two of our test instances. And
even for these two instances the difference is only minor (at most $0.6 \%$).} In
fact, the \quasikernel{} is sometimes even slightly smaller. The reason for this
is that different orders of reduction application can lead to different kernel sizes.
Table~\ref{tab:reductionStoppingcomparison} shows the effect of our
reduction tracking technique described in Section~\ref{sec:diminishing} on
\parfastker{}. It shows the algorithmic speedup \changed{over \parfastker{}}
achieved by enabling reduction tracking. \changed{We also show the relative
\quasikernel{} size increase caused by
using reduction tracking.}

\begin{table}[!th]
  \centering
  
\begin{tabular}{l|rr}%
 graph & speedup & $\Delta$ size \\
  \hline
    \begin{filecontents*}{include/reductionStoppingComparison.csv}
graph,sizeDiff,speedup
uk-2002,\numprint{+0.0} \%,\numprint{1.1}
arabic-2005,\numprint{-0.0} \%,\numprint{2.2}
gsh-2015-tpd,\numprint{-0.0} \%,\numprint{1.0}
uk-2005,\numprint{-0.0} \%,\numprint{1.0}
it-2004,\numprint{-0.0} \%,\numprint{1.8}
sk-2005,\numprint{-0.0} \%,\numprint{135.0}
uk-2007-05,\numprint{+0.1} \%,\numprint{2.4}
webbase-2001,\numprint{+0.0} \%,\numprint{1.3}
asia.osm,\numprint{-0.0} \%,\numprint{1.0}
road\_usa,\numprint{+0.0} \%,\numprint{1.0}
europe.osm,\numprint{+0.6} \%,\numprint{1.0}
rgg26,\numprint{+0.0} \%,\numprint{1.2}
rhg,\numprint{+0.0} \%,\numprint{1.0}
del24,\numprint{+0.0} \%,\numprint{1.0}
del26,\numprint{+0.0} \%,\numprint{1.0}
\end{filecontents*}
\csvreader[head to column names, late after line=\\]{include/reductionStoppingComparison.csv}{}
    {\graph & \speedup & \sizeDiff}
    \end{tabular}
\vspace*{.25cm}
  \caption{Speedup and relative
    change in kernel size change achieved by using reduction tracking.}
  \label{tab:reductionStoppingcomparison}
\end{table}
\vfill
\pagebreak
\subsection{Impact of Partitioning}

\changed{
In this section we assess the impact of the partitioning quality on
\parfastker{}. We compare ParHIP's fastest configuration (ultrafast), which we
used for our other experiments, with a higher quality  
configuration (fast) as well as a size-constrained label propagation algorithm~\cite{meyerhenke2016partitioning} (SC-LPA). SC-LPA
is a simple graph partitioning
algorithm with fast running time; however, it usually gives low quality output. We set the
imbalance for all three algorithms to $3\%$.}

\changed{
Table~\ref{tab:partitioning} shows that with ParHIP's fast configuration, the
total time for kernelization increases by a factor of $1.52$ in the geometric mean
compared to the
ultrafast configuration. The kernel size, however, remains largely unchanged. Focusing on the difference in the number of edges cut, we clearly see that ParHIP's fast configuration gives
only minimally better partitions than the ultrafast configuration: on all
instances except for \texttt{sk-2005} the difference is under $0.03 \%$. While this might be important
for certain applications, it does not seem to be worth the longer running time
for our algorithm since the quasi kernel size changes only slightly, at most $1 \%$ on most instances.
Note that
\texttt{rhg} has an empty kernel and \parfastker{} using ParHIP's ultrafast configuration finds
a \quasikernel{} of size 16, so the size difference reported in
Table~\ref{tab:partitioning} is negligible.

On the other hand, using the size constrained label propagation algorithm, the kernel size
increases drastically due to the much larger amount of cut-edges. We see that
SC-LPA produces up to $12 \%$
larger cuts than ParHIP's ultrafast configuration, resulting in \quasikernel{}s
larger by a factor of $2.4$ in the geometric mean as
more reductions are skipped because they lie on boundaries between blocks.
It is also important to note that our implementation of
SC-LPA is sequential -- it is possible that total kernelization time would be
faster with a parallel size constrained label propagation than with ParHIP's
ultrafast configuration. However, these experiments show that this simple
partitioning algorithm does not produce partitions of high enough quality to be used by
\parfastker{} -- even with faster running times -- as \quasikernel{} sizes become too large.}

\begin{table}
\centering
\begin{tabular}{l|rrr|rrr}
  &   \multicolumn{3}{c|}{ParHIP (fast)} & \multicolumn{3}{c}{SC-LPA} \\
 graph        &   $\Delta$ cut &   time & $|\mathcal{K}|$ & $\Delta$ cut &   time\footnotemark &  $|\mathcal{K}|$ \\
\hline
 uk-2002      &      -0.00 \%   &   1.56 &        1.00 &  +3.59 \%         &   4.89 &            1.80 \\
 arabic-2005  &      -0.01 \%   &   1.75 &        1.00 &  +4.77 \%         &   4.26 &            1.73 \\
 gsh-2015-tpd &      -0.00 \%   &   1.71 &        1.00 &  +6.47 \%         &   1.16 &            1.13 \\
 uk-2005      &      -0.02 \%   &   1.87 &        1.00 &  +4.68 \%         &   3.28 &            1.35 \\
 it-2004      &      -0.01 \%   &   1.36 &        1.00 &  +5.08 \%         &   1.94 &           1.51  \\
 sk-2005      &      -0.49 \%   &   1.68 &        0.97 &   *               &   *    &           *     \\
 uk-2007      &      -0.03 \%   &   1.68 &        0.99 &   *               &   *    &           *     \\
 webbase-2001 &      -0.00 \%   &   1.43 &        1.00 &  +2.84 \%         &   4.21 &            2.18 \\
 asia.osm     &      -0.00 \%   &   1.18 &        0.99 &  +9.41 \%         &   1.22 &            8.69 \\
 road\_usa     &      0.00 \%   &   1.30 &        1.00 &  +9.70 \%         &   1.54 &            5.94 \\
 europe.osm   &       0.00 \%   &   1.14 &        0.99 &  +8.71 \%         &   1.21 &           33.02 \\
 rgg26          &    -0.01 \%   &   1.20 &        1.00 & +10.79 \%         &   2.84 &            1.02 \\
 rhg          &      -0.00 \%   &   1.94 &        0.62 &   *               &   *    &            *    \\
 del24 &             -0.01 \%   &   1.70 &        1.00 & +12.27 \%         &   1.11 &            1.16 \\
 del26        &      -0.01 \%   &   1.64 &        1.00 & +12.30 \%         &   1.27 &            1.16 \\

\end{tabular}
\vspace*{.25cm}
\caption{Comparison of \parfastker{'s} running time and \quasikernel{} size with the ultrafast configuration
  of ParHIP to the fast configuration and a size constrained label propagation
  algorithm (SC-LPA). Times and kernel sizes are divided by the respective value for the
  ultrafast configuration. The column `$\Delta$ cut' gives the difference in the
  number edges cut by the partition divided by the total number of edges in the
  graph (in comparison to ParHIP's ultrafast configuration).}
\label{tab:partitioning}
\end{table}

\begin{figure*}[t]
  \centering
  \begin{subfigure}[t]{0.32\textwidth}
    \includegraphics[width=\textwidth]{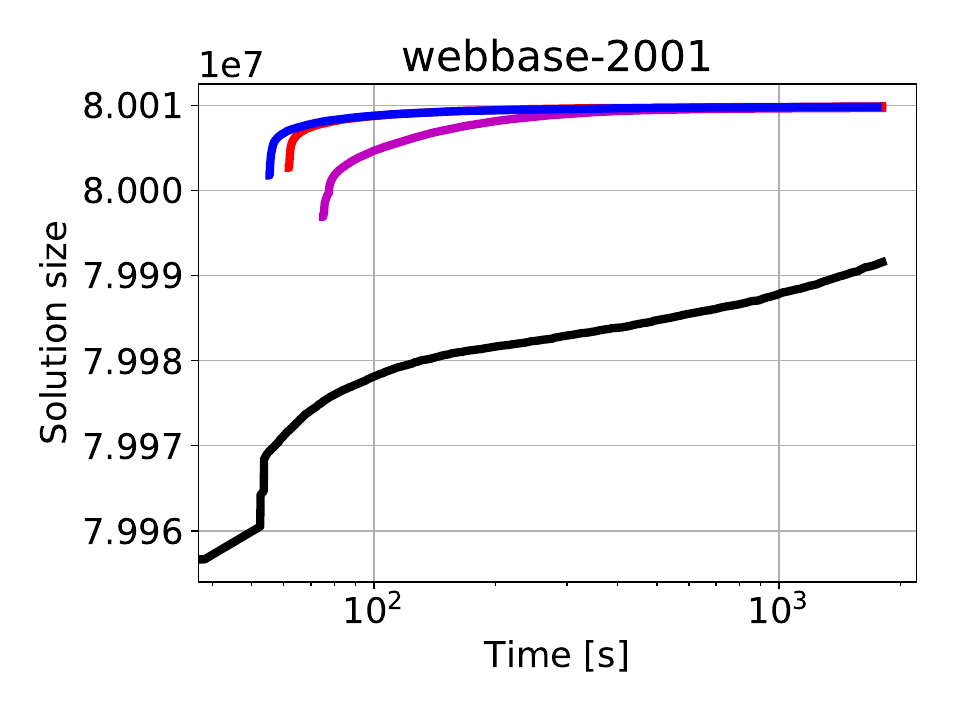}
  \end{subfigure}
  \begin{subfigure}[t]{0.32\textwidth}
    \includegraphics[width=\textwidth]{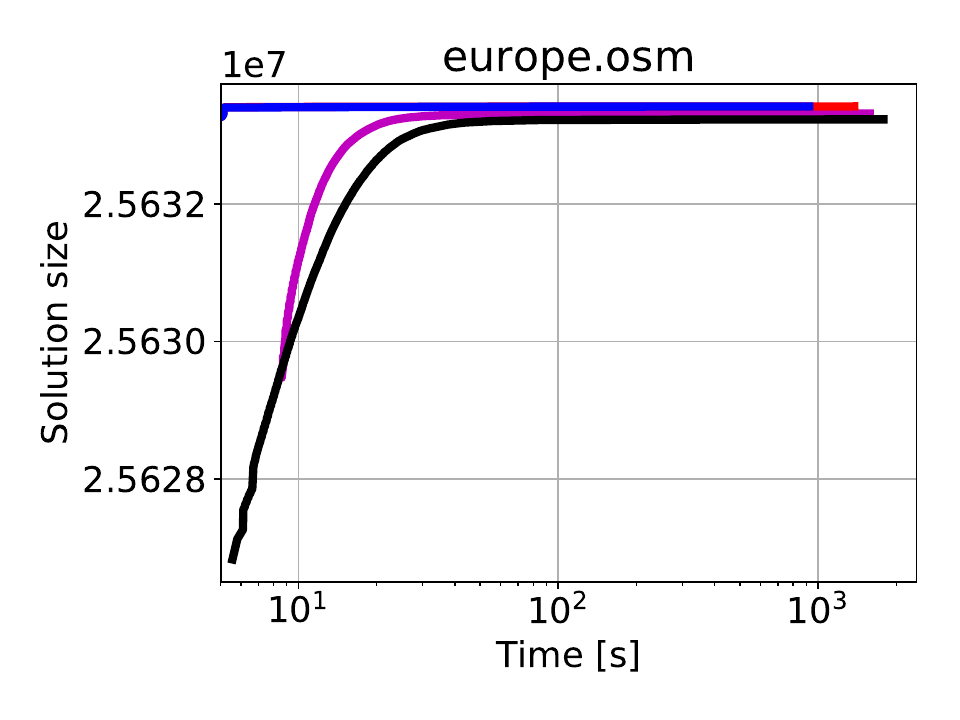}
  \end{subfigure}
  \begin{subfigure}[t]{0.32\textwidth}
    \includegraphics[width=\textwidth]{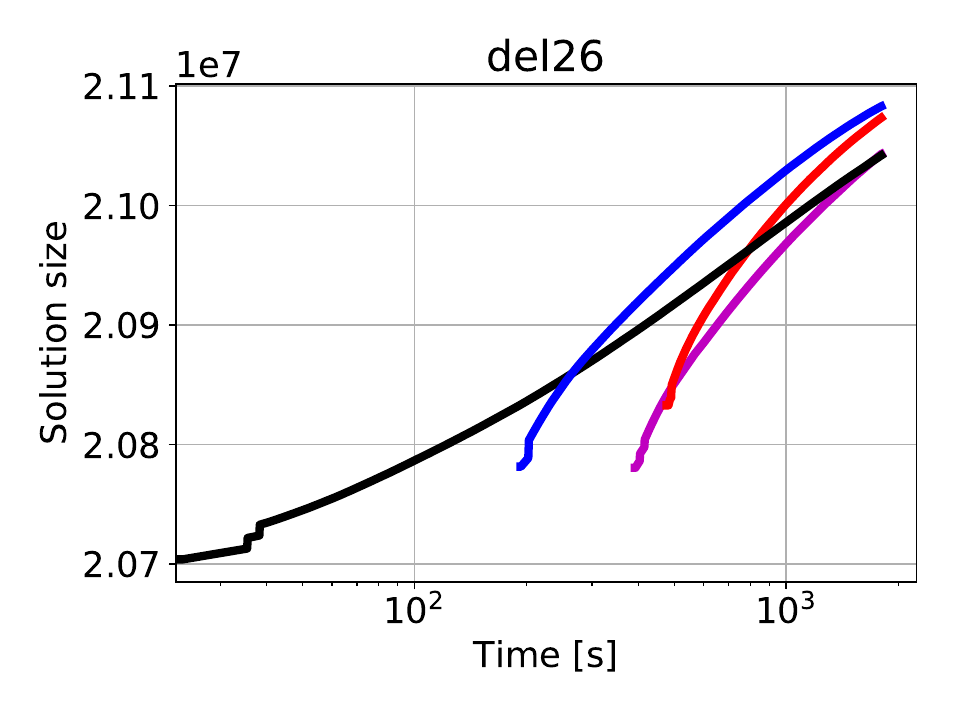}
  \end{subfigure}
  \begin{subfigure}[t]{0.32\textwidth}
    \includegraphics[width=\textwidth]{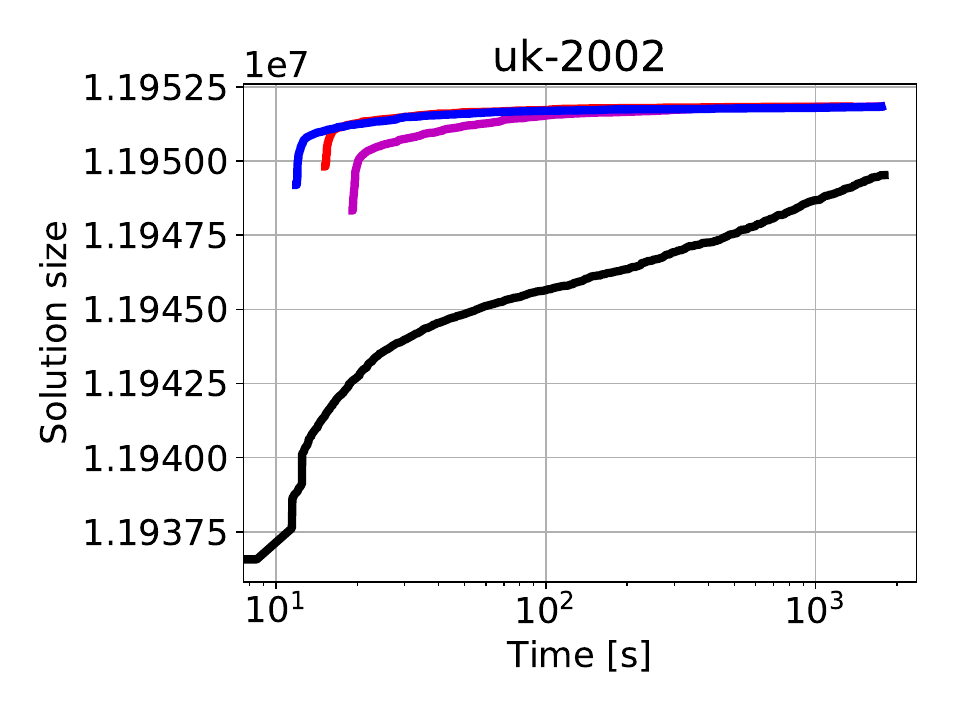}
  \end{subfigure}
  \begin{subfigure}[t]{0.32\textwidth}
    \includegraphics[width=\textwidth]{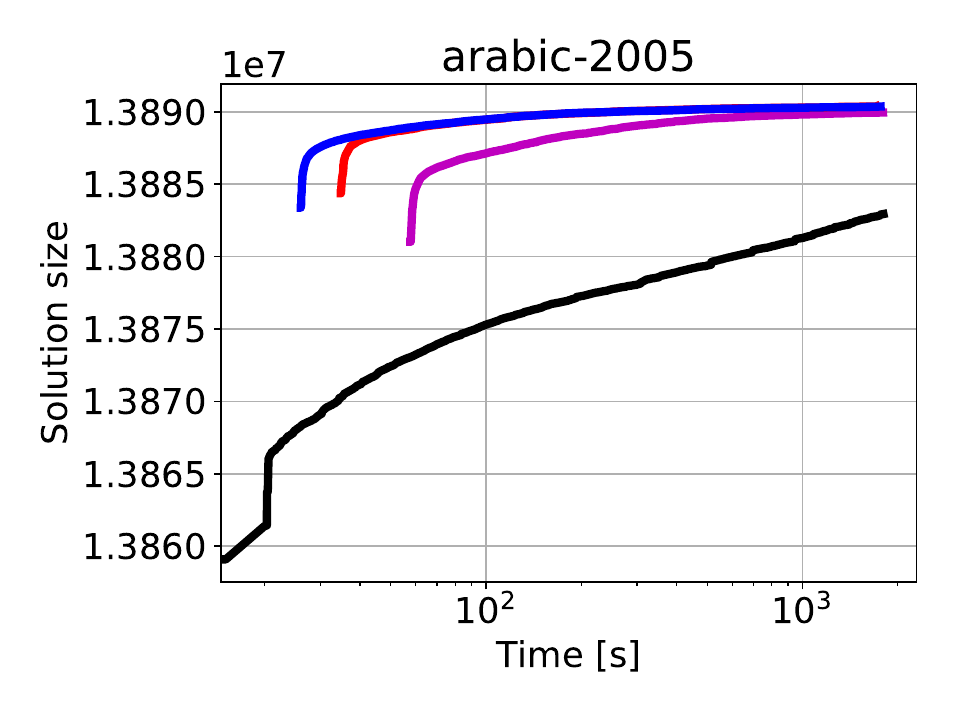}
  \end{subfigure}
  \begin{subfigure}[t]{0.32\textwidth}
    \includegraphics[width=\textwidth]{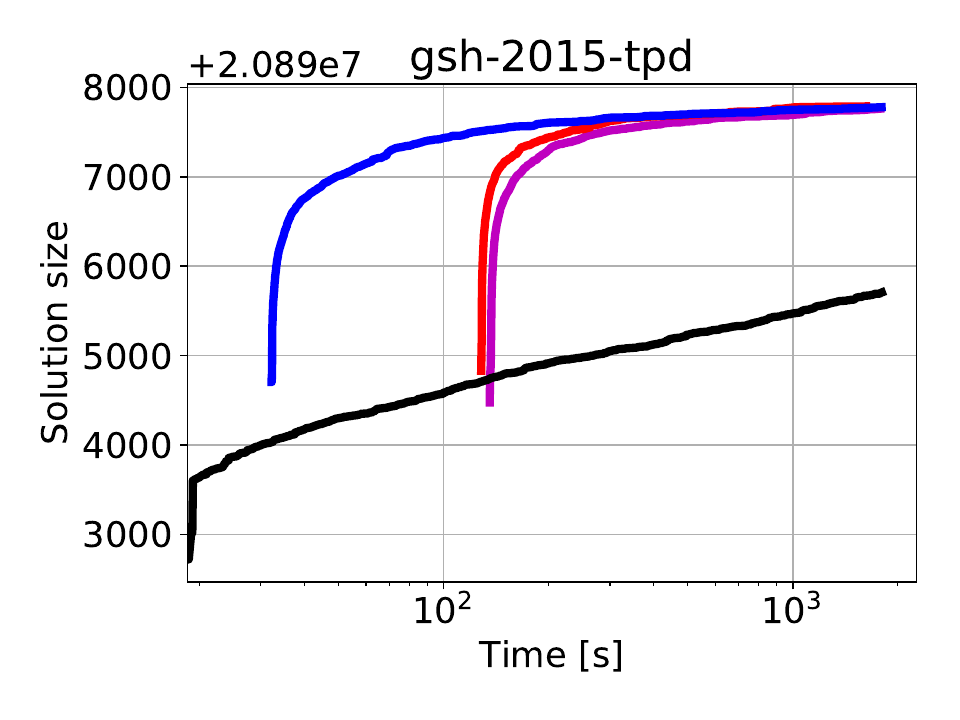}
  \end{subfigure}
  \begin{subfigure}[t]{0.32\textwidth}
    \includegraphics[width=\textwidth]{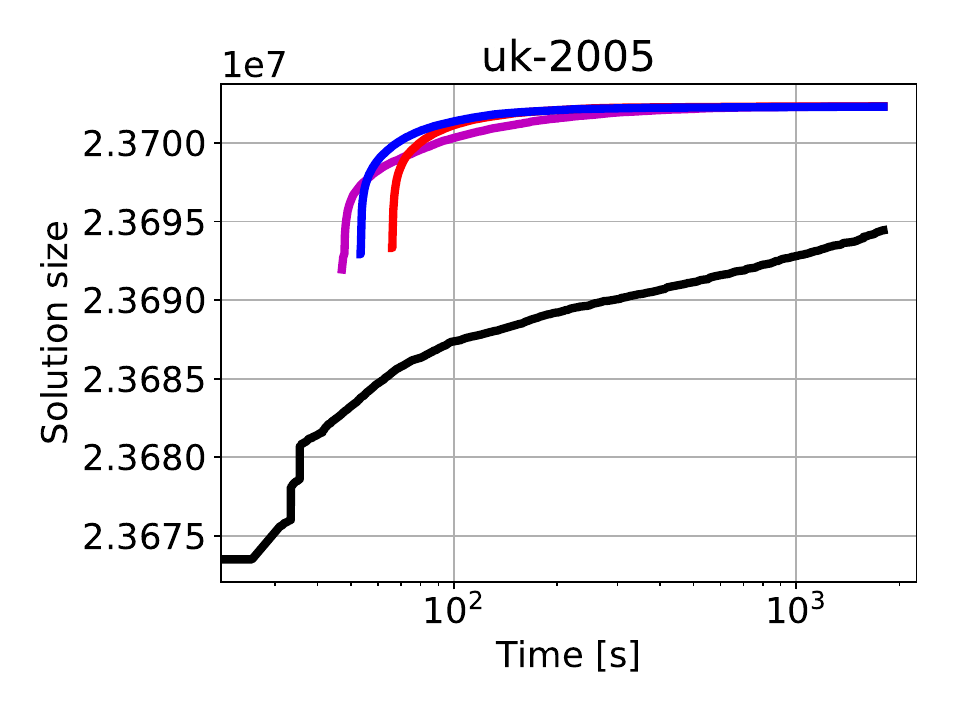}
  \end{subfigure}
  \begin{subfigure}[t]{0.32\textwidth}
    \includegraphics[width=\textwidth]{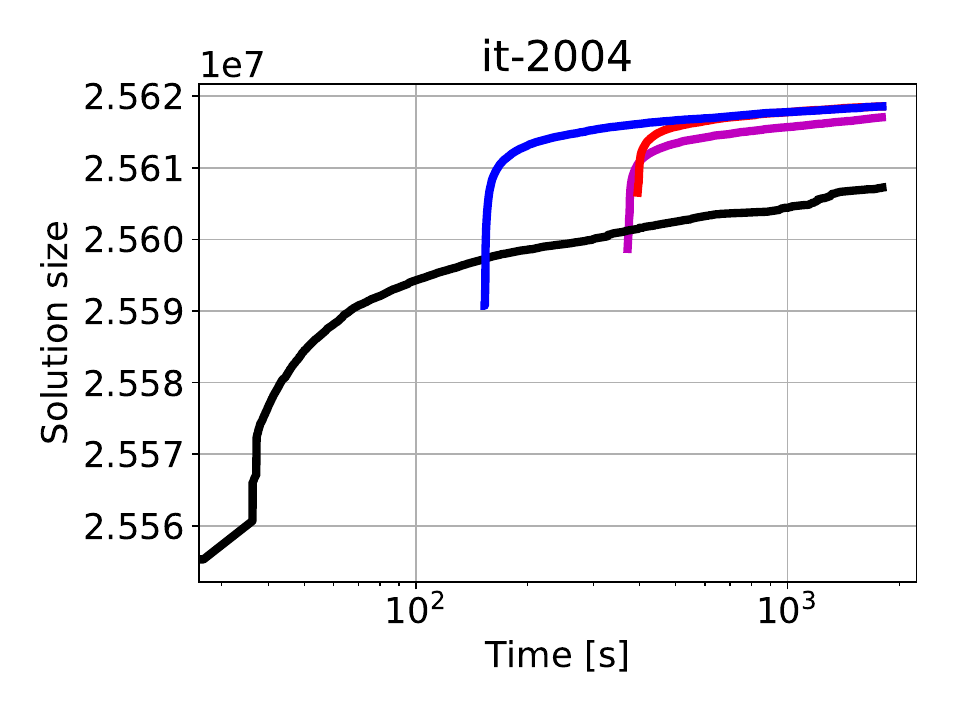}
  \end{subfigure}
  \begin{subfigure}[t]{0.32\textwidth}
    \includegraphics[width=\textwidth]{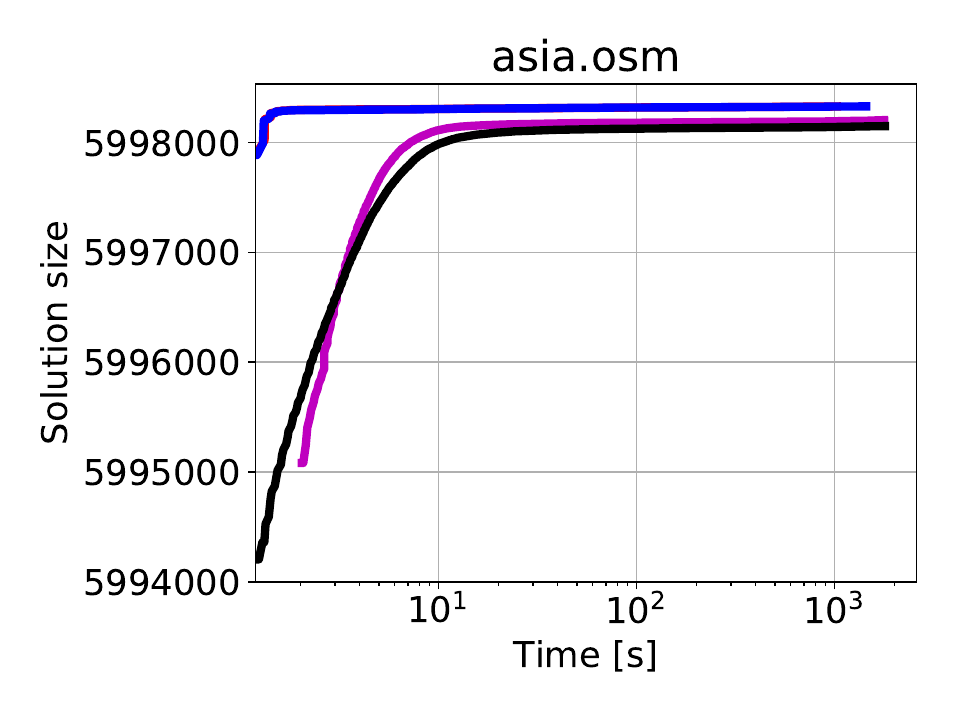}
  \end{subfigure}
  \begin{subfigure}[t]{0.32\textwidth}
    \includegraphics[width=\textwidth]{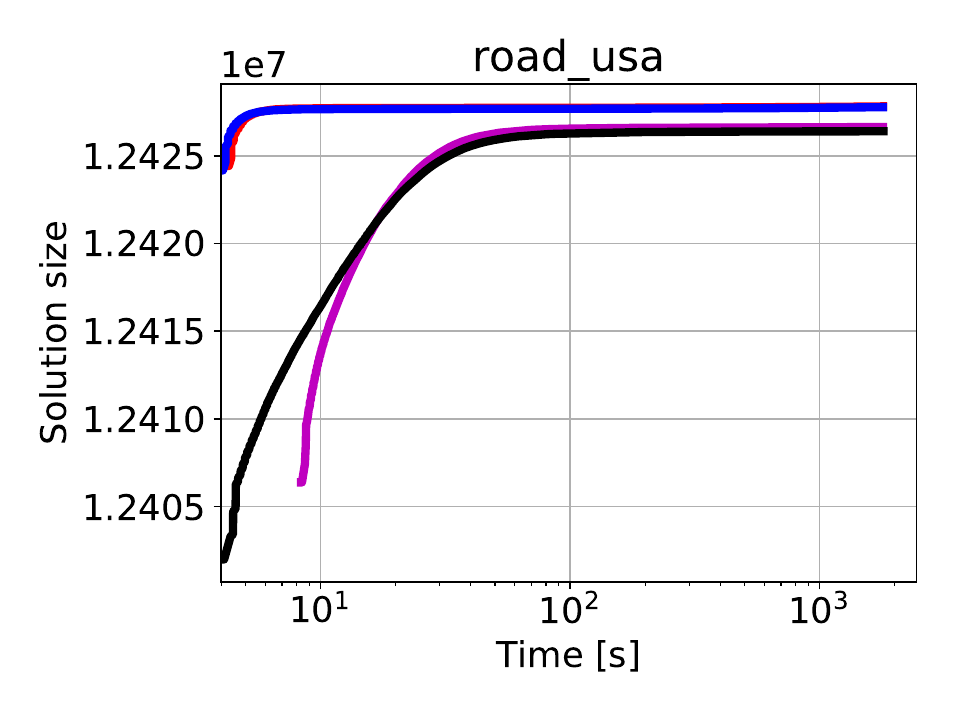}
  \end{subfigure}
  \begin{subfigure}[t]{0.32\textwidth}
    \includegraphics[width=\textwidth]{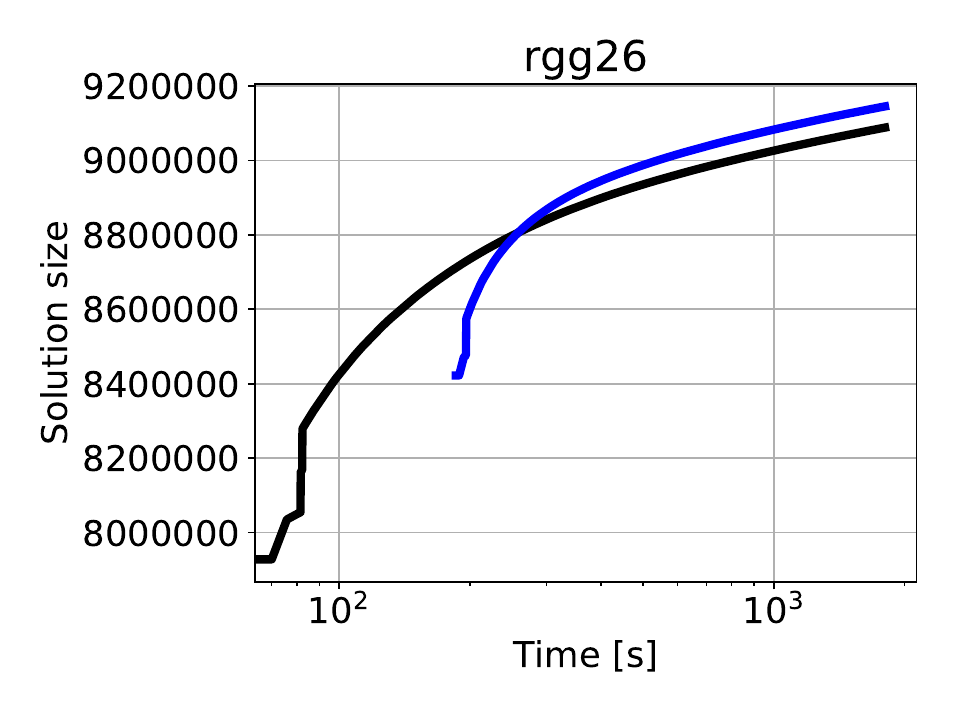}
  \end{subfigure}
  \begin{subfigure}[t]{0.32\textwidth}
    \includegraphics[width=\textwidth]{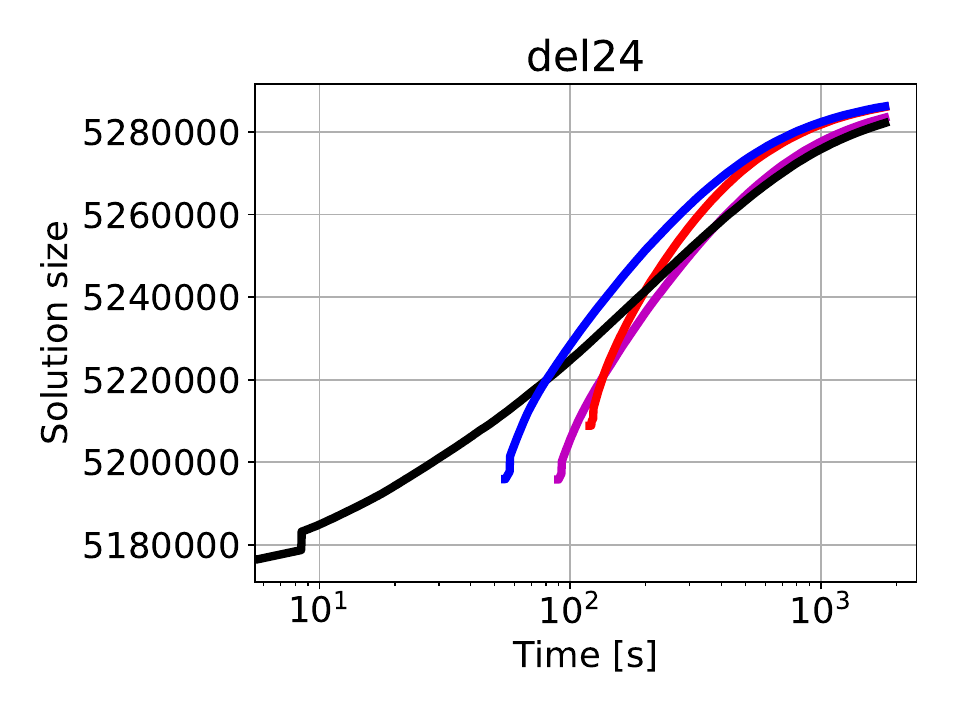}
  \end{subfigure}
  \begin{subfigure}[t]{\textwidth}
    \includegraphics[width=\textwidth]{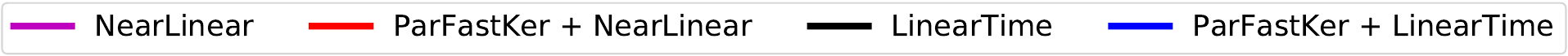}
  \end{subfigure}
  \caption{Solution size over time of \lineartime{} and \nearlinear{} in the
    original version and with \parfastker{} as preprocessing step
    {(marked with "ParFastKer +")}. On
    \texttt{rgg26}, \nearlinear{} did not find an initial solution within the
    time limit. {Note that by \lineartime{} and \nearlinear{} we refer to the
    full local search algorithms, not just their kernelization parts.}}
  \label{fig:local_search}
\end{figure*}
 
\subsection{Local Search on the Quasi Kernel}
We now demonstrate the impact that quickly finding a small \quasikernel{}
has on algorithms for finding large independent sets with local search.
Currently, the algorithms with the best trade-off between speed and solution quality are
\lineartime{} and \nearlinear{} by
Chang~\etal~\cite{chang2017computing}. In our previous experiments, we
compared only against the \lineartime{} and \nearlinear{} kernelization. However, we now run the
full algorithm of Chang~\etal, which first kernelizes the graph, then invokes ``reducing-peeling'' to compute
an initial solution for local search, and then runs local search~\cite{andrade-2012} on the kernel.
We compare their original algorithms against variants that first kernelize the graph with \parfastker{} 
and then run on the \quasikernel{}. We use a time limit of 30 minutes, including kernelization and finding an initial solution. Figure~\ref{fig:local_search}
shows the size of the independent set found over time for the largest web graph,
road network and generated graph from our benchmark set (excluding graphs that
cannot be processed due to the original 32-bit implementation of \lineartime{} and \nearlinear).
We ran local search three times using
a different input seed for each run; however, we use the same input graph (either the
  original graph or a \quasikernel{} found by \parfastker{}) for each run. The plots show at any given time the geometric mean of the current best
  solution of all runs.

\footnotetext{\changed{The implementation we use for SC-LPA is a 32-bit implementation. Graphs that
    cannot be processed by it are marked with a star (*).}}
For web graphs we see that using \parfastker{}'s
\quasikernel{}, the independent set found is much larger (\numprint{80009858}
for \texttt{webbase-2001}) than the one found by
\lineartime{} (\numprint{18286} vertices less) and \nearlinear{} only converges
to approximately the size found using the
\quasikernel{} after several hundred seconds. On road networks, we observe interesting
behavior: local search seemingly converges for all algorithms, but to \emph{different} independent
set sizes: the smaller the initial kernel size, the larger independent set size.
On \texttt{europe.osm}, the final solution size is
\numprint{25633238} for \lineartime{}, \numprint{84} more for \nearlinear{} and
\numprint{188} more for both versions that use \parfastker{}'s \quasikernel{}. Also, using \parfastker{}'s \quasikernel{}, the algorithm converges
much faster. 
In particular, after an initial improvement over the starting solution,
that takes about \numprint{0.1} seconds plus the time for kernelization and finding an initial
solution (which is about \numprint{5} seconds),
very few changes occur with \parfastker{}'s \quasikernel{}.
\lineartime{} and \nearlinear{}, on the other hand, make an increase of several hundred vertices
for the first \numprint{30} to \numprint{40} seconds. 
On the Delaunay triangulation graphs, the smaller
\quasikernel{} enables local search to find larger independent sets. 
\subsection{Improving High Quality Heuristic Algorithms}
\begin{figure*}[t]
  \centering
  \begin{subfigure}[t]{0.32\textwidth}
    \includegraphics[width=\textwidth]{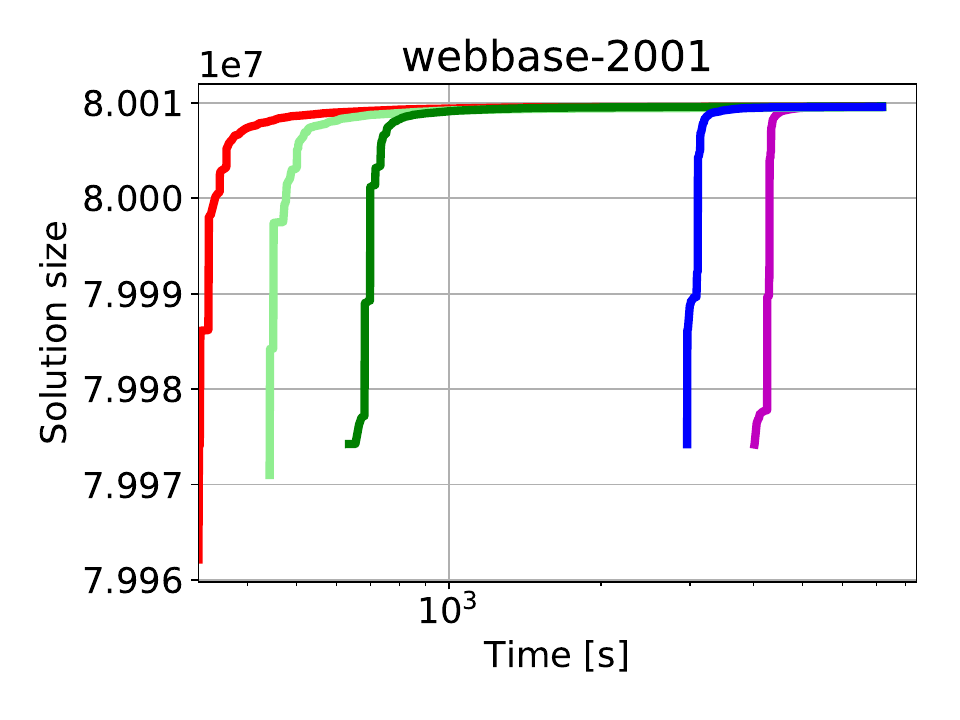}
  \end{subfigure}
  \begin{subfigure}[t]{0.32\textwidth}
    \includegraphics[width=\textwidth]{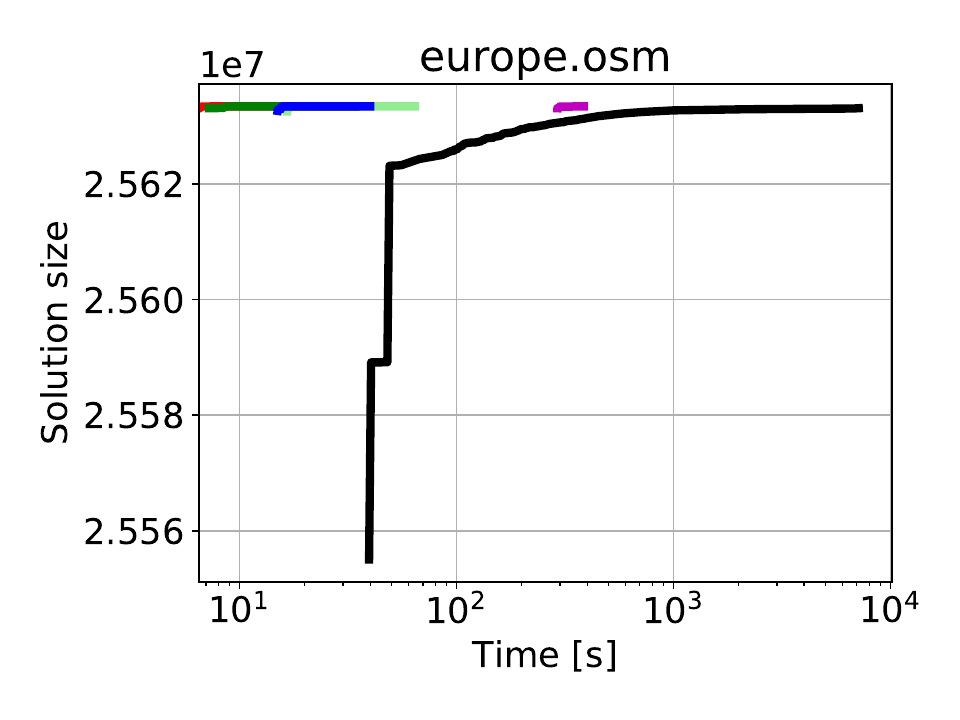}
  \end{subfigure}
  \begin{subfigure}[t]{0.32\textwidth}
    \includegraphics[width=\textwidth]{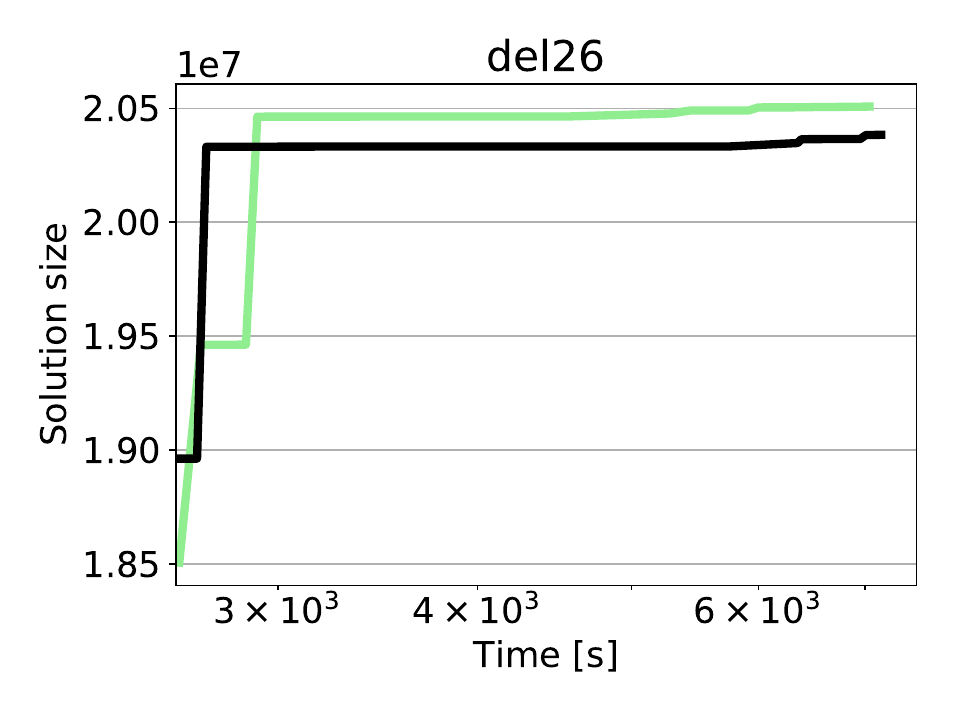}
  \end{subfigure}
  \begin{subfigure}[t]{0.32\textwidth}
    \includegraphics[width=\textwidth]{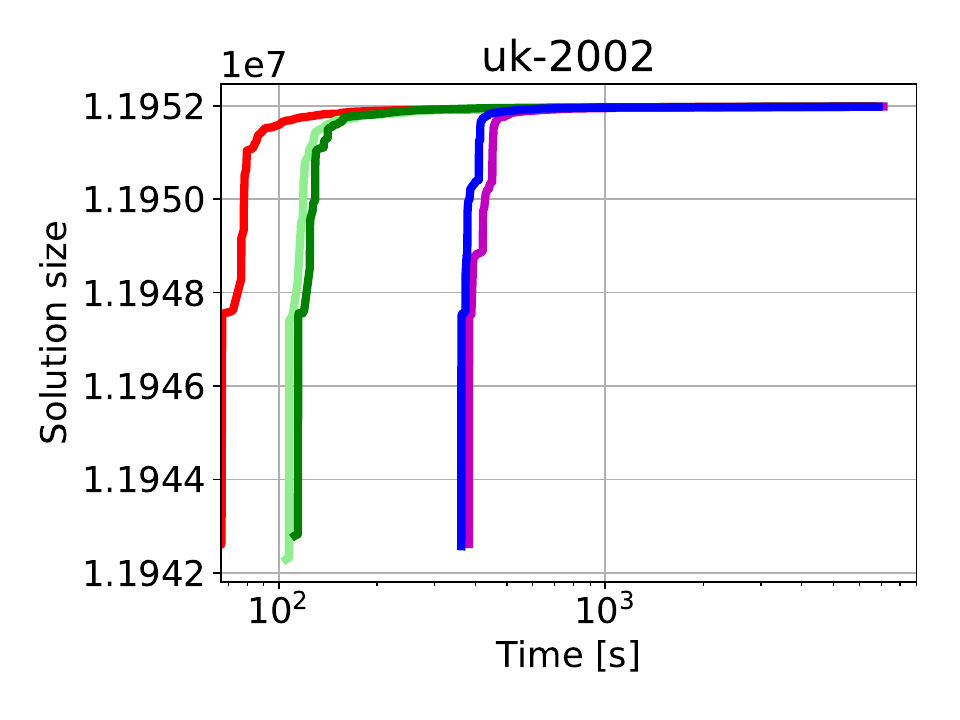}
  \end{subfigure}
  \begin{subfigure}[t]{0.32\textwidth}
    \includegraphics[width=\textwidth]{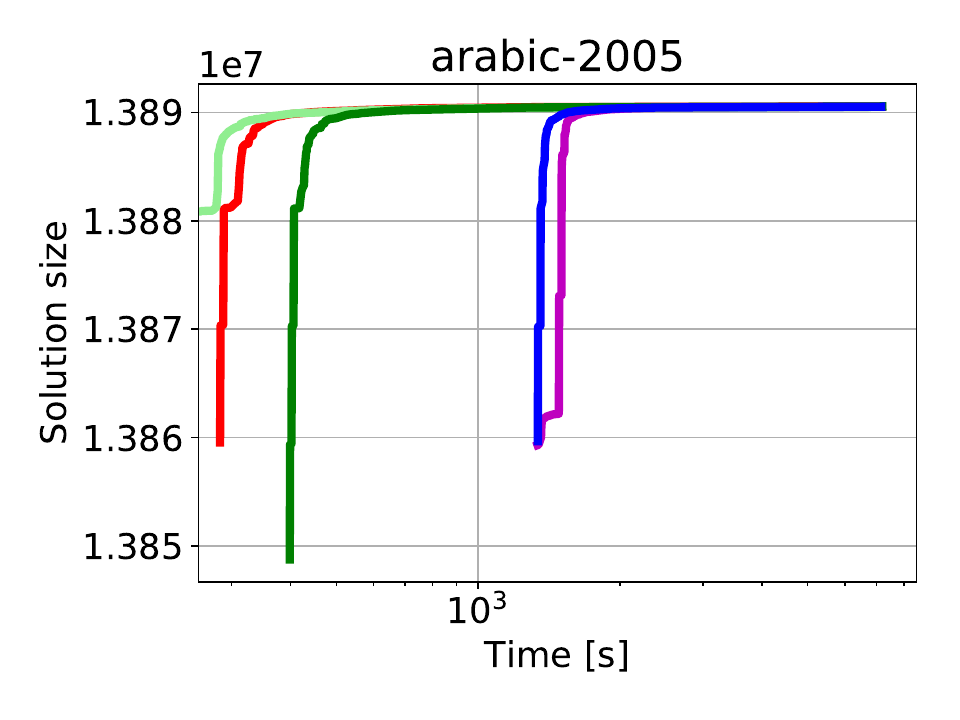}
  \end{subfigure}
  \begin{subfigure}[t]{0.32\textwidth}
    \includegraphics[width=\textwidth]{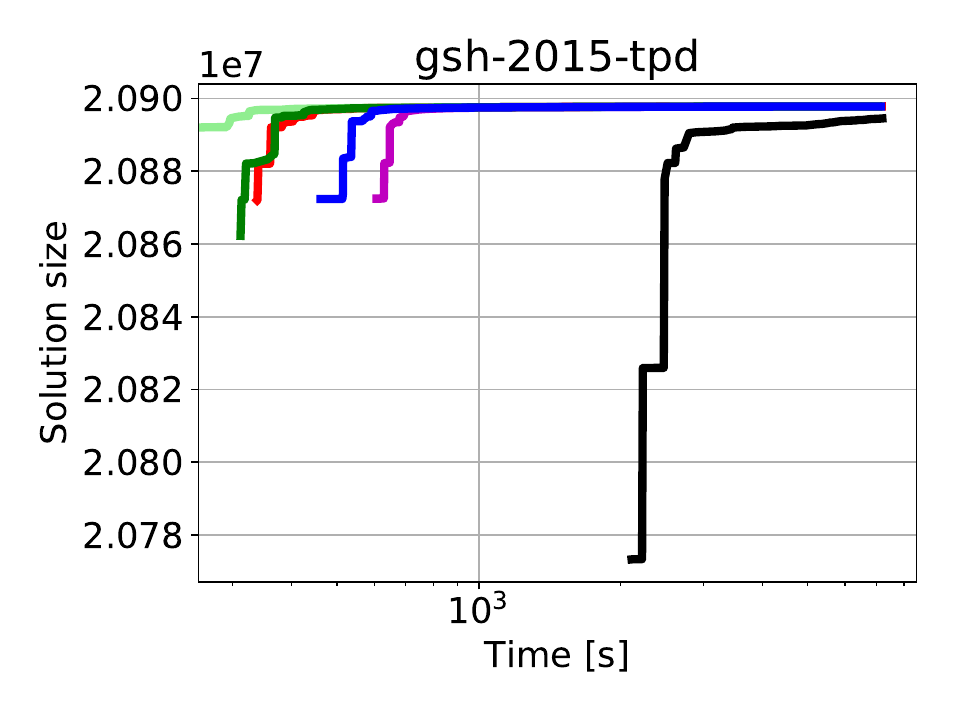}
  \end{subfigure}
  \begin{subfigure}[t]{0.32\textwidth}
    \includegraphics[width=\textwidth]{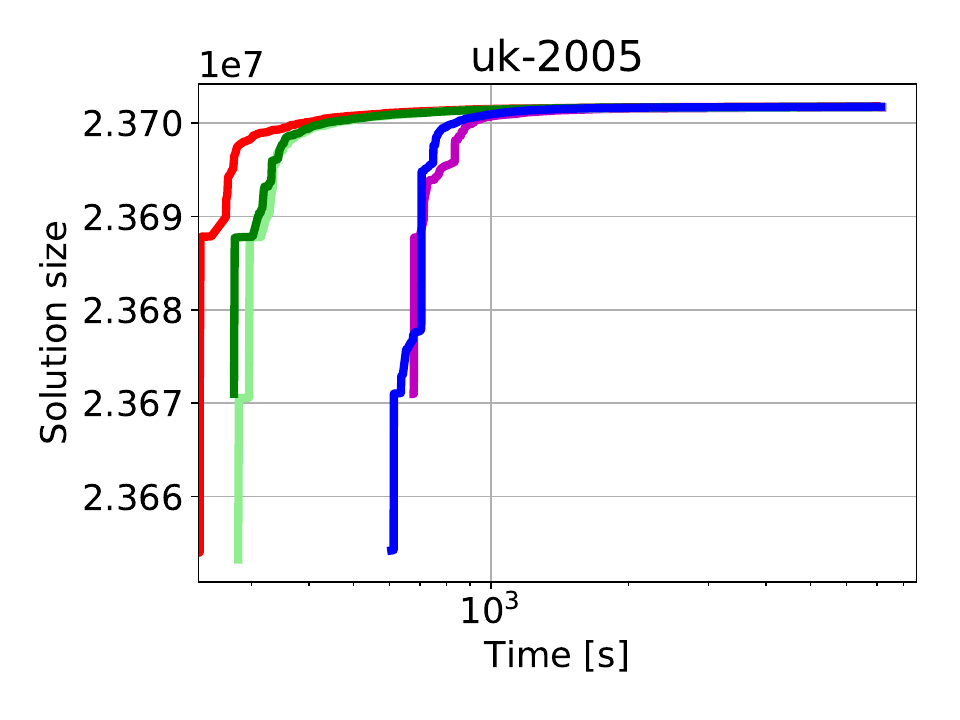}
  \end{subfigure}
  \begin{subfigure}[t]{0.32\textwidth}
    \includegraphics[width=\textwidth]{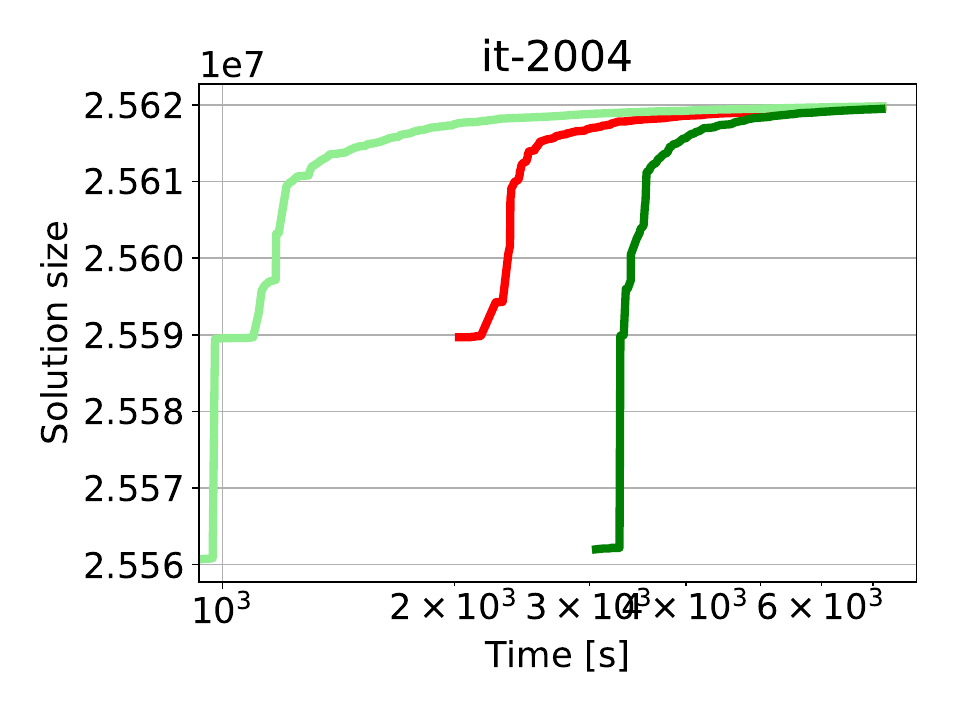}
  \end{subfigure}
  \begin{subfigure}[t]{0.32\textwidth}
    \includegraphics[width=\textwidth]{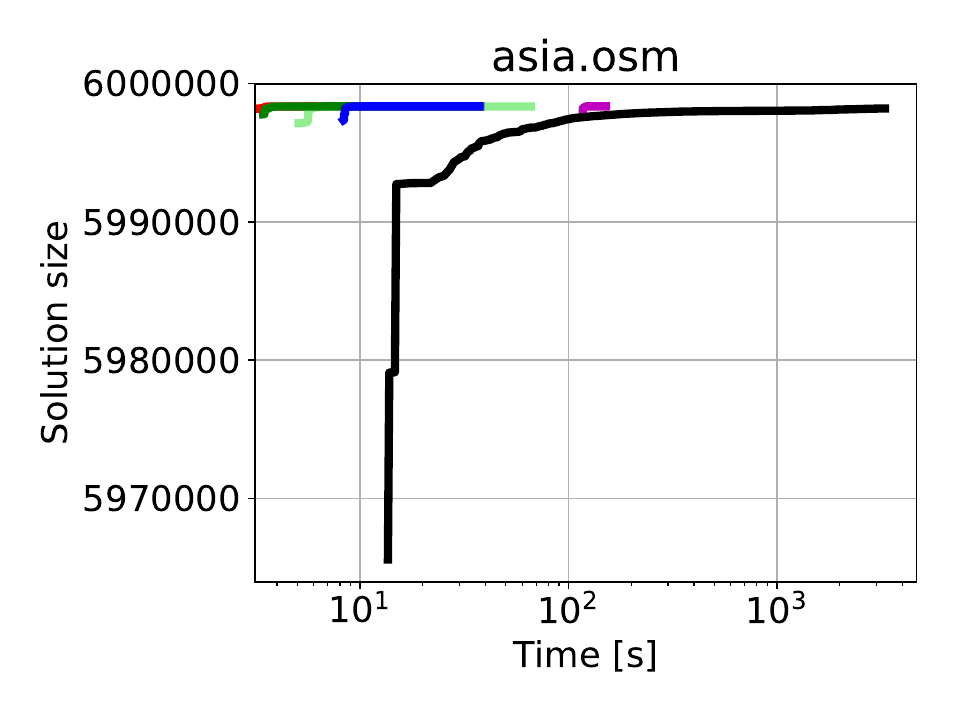}
  \end{subfigure}
  \begin{subfigure}[t]{0.32\textwidth}
    \includegraphics[width=\textwidth]{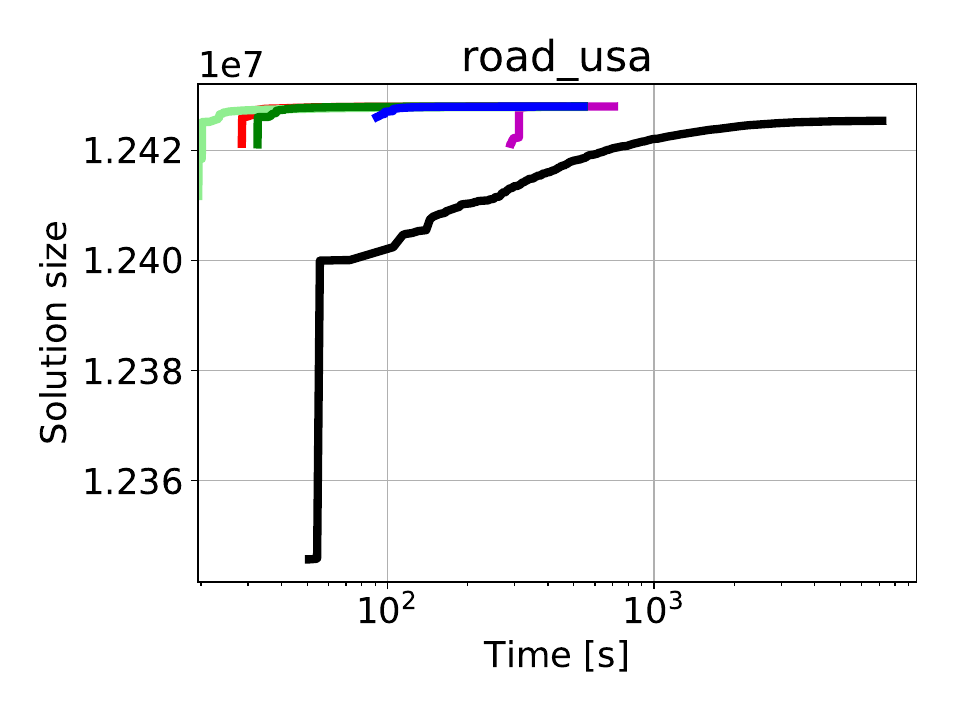}
  \end{subfigure}
  \begin{subfigure}[t]{0.32\textwidth}
    \includegraphics[width=\textwidth]{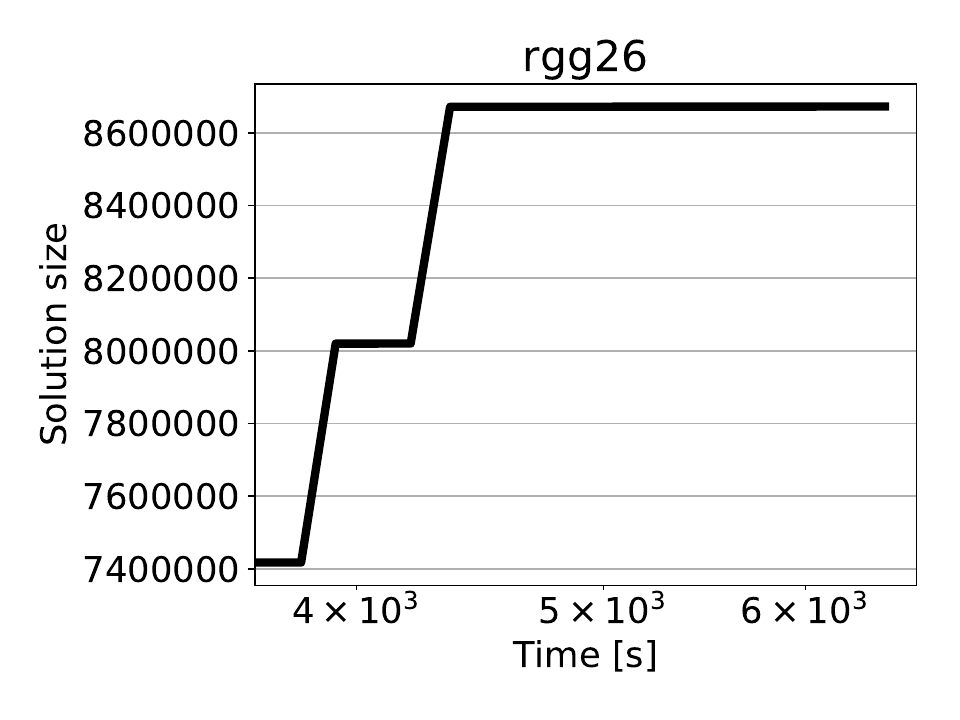}
  \end{subfigure}
  \begin{subfigure}[t]{0.32\textwidth}
    \includegraphics[width=\textwidth]{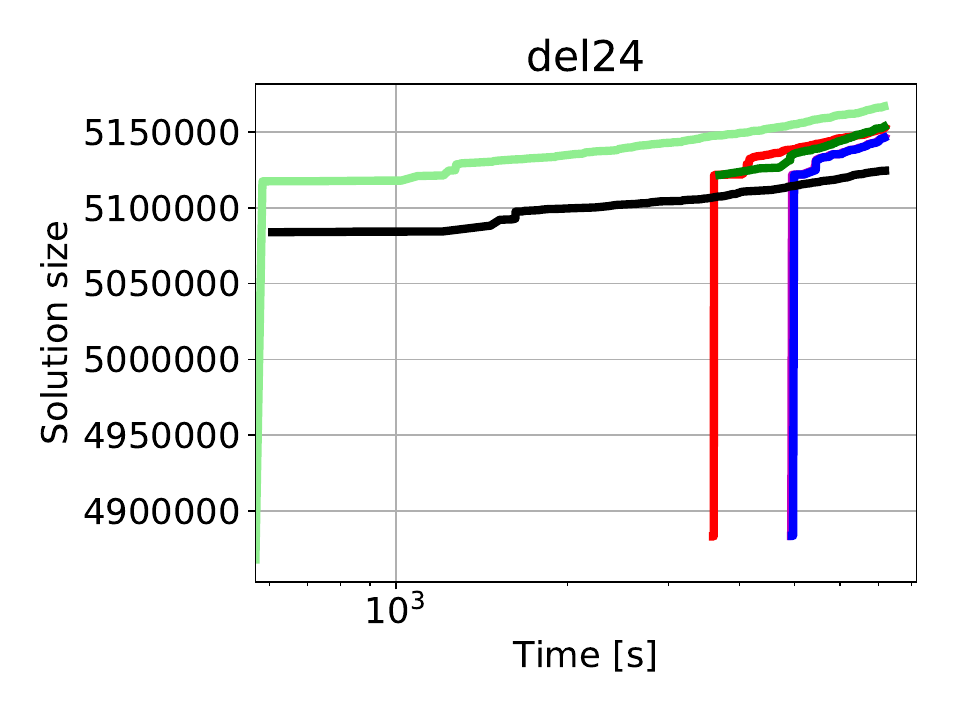}
  \end{subfigure}
  \begin{subfigure}[t]{\textwidth}
    \includegraphics[width=\textwidth]{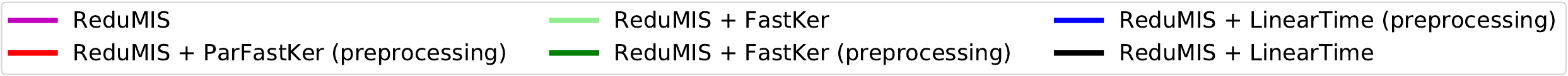}
  \end{subfigure}
  \caption{\changed{Experimental results for different variations of \redumis{}. Variants
    marked with (preprocessing) use the indicated kernelization algorithm as
    preprocessing and then run the original \redumis{} algorithm on the resulting
    kernel. The other variants replace the kernelization algorithm used inside
    of \redumis{}. All runs were done with a time limit of two hours.}}
  \label{fig:redumis}
\end{figure*}

\changed{
We now show the impact that our fast kernelization makes on a heuristic
algorithm that is tailored towards very high quality solutions. In particular,
we consider \redumis{} by Lamm~\etal~\cite{lamm2017finding}. \redumis{} first finds the
kernel of the input graph using \vcsolver{}'s kernelization algorithm and then uses an evolutionary algorithm to find a
large independent set of the kernel. It then fixes the $10\%$ lowest degree
vertices from the independent set into the solution, removing them and their
neighborhood. After removal, the graph has changed and kernelization can be run again. This is
repeated until a time limit is met or the graph has been fully reduced. We show
results for two different experiments: Replacing the kernelization algorithm used
by \redumis{} internally by integrating a different kernelization algorithm, and first reducing the input graph using different
kernelization algorithms as a preprocessing step and then using the resulting kernel as input to
the original version of \redumis{} which uses \vcsolver{} for kernelization. The
time limit for all experiments in this section was set to two
hours. Figure~\ref{fig:redumis} shows the results of these experiments.
}

\changed{
We see that for many instances, the versions with \fastker{} and \parfastker{}
outperform the other versions: The algorithm starts finding solutions earlier
and thus has more time to improve its solution until the time limit is
met. On some graphs the version that integrates \fastker{} into \redumis{} performs
significantly better than the versions that use our algorithms as a
preprocessing step: on 6 out of 12 instances \redumis{} + \fastker{} performs best
among all other variants and on 5 out of 12 instances \redumis{} + \parfastker{}
(preprocessing) outperforms the other variants -- only slightly in some cases, though.
The advantage of the integrated version is especially noticeable on \texttt{del26}, where the
non-integrated versions fail to find a solution within the time limit, and
\texttt{it-2004} and \texttt{del24}, where the non-integrated versions start
finding solutions much later than the integrated version and thus have less time
to improve their initial solution. Possible explanations for this are: \vcsolver{} (which is used
by \redumis{} in all versions that just use a different kernelization algorithm as
preprocessing step) is slow in applying the remaining reductions that \fastker{}
and \parfastker{} did not apply, or there are large graph size reductions in
later stages of the algorithm which can be sped up in the integrated version. As
on most graphs, the plots for the integrated \fastker{} version and the
preprocessing versions behave very similar after finding a first solution, we
assume the former to be the case.
}

\changed{
The version with integrated \lineartime{} kernelization cannot reduce the graph
enough for \redumis{} to find any solution on the graphs with high degree vertices
(road networks as well as the geometric graph instances shown here usually do
not have high degree vertices).
}

\section{Conclusion}
We presented an efficient parallel kernelization algorithm based on graph partitioning and parallel bipartite maximum matching, vertex pruning as well as reduction tracking. 
On the one hand, our algorithm produces kernels that are orders of magnitude smaller than the fastest kernelization methods, while having a similar execution time. 
On the other hand, our algorithm is able to compute kernels with size comparable to the smallest known kernels, but up to two orders of magnitude faster that previously possible.
Experiments with local search algorithms show that we find larger independent sets faster.
In future work, \changed{we want to parallelize the \lineartime{} algorithm by
  Chang~\etal~\cite{chang2017computing} so that our algorithm is fully
parallel}, apply our parallel kernelization techniques in
more MIS algorithms, such as exact branch-and-reduce~\cite{akiba-tcs-2016}, {explore techniques for further parallelizing the LP reduction}, and transfer our techniques to other problems that use kernelization~\changed{\cite{Dehne2007,jansen2013data,guo2007invitation,etscheid2018linear}}. 

\textbf{Acknowledgements.} Our research was supported by the Gottfried Wilhelm Leibniz Prize 2012, the Large-Scale Data Management and Analysis (LSDMA) project in the Helmholtz Association, has received funding from the European Research Council under the European Union's Seventh Framework Programme (FP/2007-2013) / ERC Grant Agreement no. 340506 and was supported by DFG grant SCHU 2567/1-2.
\bibliographystyle{plain}
\bibliography{references}

\newpage
\appendix

\section{Detailed Results}
\label{app:detailed_results}

Here, we provide detailed results of our experiments. In addition to the
time to reach a \quasikernel{}, we also provide the time it takes to reach a
full kernel. We do this by first applying our algorithm as described throughout the
paper to find a \quasikernel. We then apply the remaining reductions by running
sequentially and disabling the inexact reduction pruning technique described in
Section~\ref{sec:diminishing}.

In the comparisons to \vcsolver{}, we also provide columns for a ``same
size comparison''. This is found by logging the current time and size throughout
the algorithms. When comparing two algorithms with different kernel sizes, the
time column of the same size comparison then reports the first time stamp at
which the algorithm with the smaller kernel size logged a size smaller than (or
equal to) the final size of the algorithm with the larger kernel.

The implementation by Chang~\etal{} uses 32-bit integers as edge identifiers, so
they cannot process graphs with $2^{32}\approx4.29$B or more edges. Respective entries in
the tables are marked with a star~(*). As our algorithm uses their \lineartime{}
implementation as a preprocessing step, for graphs with too many edges, we use
the original graph as input to our algorithm instead of the kernel found by \lineartime{}.
\begin{figure*}[!h]
  \centering
\footnotesize
\begin{tabular}{l|rrrr|rrr}%
  & \multicolumn{4}{c|}{\quasikernel} & \multicolumn{3}{c}{kernel}\\
 graph & $|\mathcal{K}|$ & \lineartime{} [s] & all reductions [s] & total [s] & $|\mathcal{K}|$ &  time [s] & total [s] \\
  \hline
    \begin{filecontents*}{include/ourAlgorithmSequential.csv}
graph,ourSize,LinearTimeTime,quasiKernelTime,quasiKernelTotal,kernelSize,kernelTime,kerneltotal
uk-2002,\numprint{255497},\numprint{1.5},\numprint{58.6},\numprint{60.1},\numprint{255497},\numprint{1.6},\numprint{61.7}
arabic-2005,\numprint{610715},\numprint{2.6},\numprint{145.5},\numprint{148.0},\numprint{610697},\numprint{4.4},\numprint{152.5}
gsh-2015-tpd,\numprint{425645},\numprint{11.6},\numprint{54.8},\numprint{66.4},\numprint{425645},\numprint{1.8},\numprint{68.2}
uk-2005,\numprint{854511},\numprint{2.5},\numprint{129.3},\numprint{131.9},\numprint{854511},\numprint{3.7},\numprint{135.6}
it-2004,\numprint{1651630},\numprint{3.3},\numprint{496.4},\numprint{499.7},\numprint{1645591},\numprint{741.7},\numprint{1241.5}
sk-2005,\numprint{3265615},*,\numprint{2349.8},\numprint{2349.8},\numprint{3256645},\numprint{793.0},\numprint{3142.8}
uk-2007-05,\numprint{3631546},*,\numprint{2073.4},\numprint{2073.4},\numprint{3627912},\numprint{2830.7},\numprint{4904.1}
webbase-2001,\numprint{821492},\numprint{13.0},\numprint{277.8},\numprint{290.8},\numprint{821092},\numprint{59.2},\numprint{350.0}
asia.osm,\numprint{34930},\numprint{0.8},\numprint{0.8},\numprint{1.6},\numprint{34930},\numprint{0.1},\numprint{1.6}
road\_usa,\numprint{247395},\numprint{2.5},\numprint{5.4},\numprint{8.0},\numprint{247395},\numprint{0.4},\numprint{8.3}
europe.osm,\numprint{14066},\numprint{4.1},\numprint{1.7},\numprint{5.8},\numprint{14066},\numprint{0.1},\numprint{5.9}
rgg26,\numprint{49843887},\numprint{1.0},\numprint{13571.6},\numprint{13572.6},\numprint{49838878},\numprint{1024.9},\numprint{14597.6}
rhg,\numprint{0},*,\numprint{164.5},\numprint{164.5},\numprint{0},\numprint{11.3},\numprint{175.8}
del24,\numprint{12884514},\numprint{0.2},\numprint{141.8},\numprint{142.0},\numprint{12877164},\numprint{173.4},\numprint{315.4}
del26,\numprint{51701698},\numprint{0.7},\numprint{718.2},\numprint{718.9},\numprint{51624241},\numprint{708.3},\numprint{1427.2}
\end{filecontents*}
\csvreader[head to column names, late after line=\\]{include/ourAlgorithmSequential.csv}{}
    {\graph & \ourSize & \LinearTimeTime & \quasiKernelTime & \quasiKernelTotal & \kernelSize & \kernelTime & \kerneltotal}
    \end{tabular}

    \caption{Kernel sizes and kernelization times for \fastker{} to reach a
      \quasikernel{} and to reach a full kernel by running our algorithm without
      stopping the reduction application on the \quasikernel.}
\end{figure*}

\begin{figure*}[!h]
  \centering
\footnotesize
\begin{tabular}{l|rrrrr|rrr}%
  & \multicolumn{5}{c|}{\quasikernel} & \multicolumn{3}{c}{kernel}\\
 graph & $|\mathcal{K}|$ & \lineartime{} [s] & part. [s] & all reductions [s] & total [s] & $|\mathcal{K}|$ &  time [s] & total [s] \\
  \hline
    \begin{filecontents*}{include/ourAlgorithmParallel.csv}
graph,ourSize,LinearTimeTime,partitioningTime,quasiKernelTime,quasiKernelTotal,kernelSize,kernelTime,kerneltotal
uk-2002,\numprint{266328},\numprint{1.5},\numprint{4.9},\numprint{5.4},\numprint{11.8},\numprint{255594},\numprint{4.8},\numprint{16.6}
arabic-2005,\numprint{628850},\numprint{2.6},\numprint{8.3},\numprint{14.8},\numprint{25.7},\numprint{610288},\numprint{21.0},\numprint{46.6}
gsh-2015-tpd,\numprint{486328},\numprint{11.6},\numprint{13.5},\numprint{6.6},\numprint{31.7},\numprint{425751},\numprint{19.7},\numprint{51.5}
uk-2005,\numprint{901896},\numprint{2.5},\numprint{35.5},\numprint{14.8},\numprint{52.8},\numprint{854383},\numprint{18.1},\numprint{70.9}
it-2004,\numprint{1697934},\numprint{3.3},\numprint{30.0},\numprint{117.8},\numprint{151.1},\numprint{1645643},\numprint{148.2},\numprint{299.3}
sk-2005,\numprint{3504786},*,\numprint{70.0},\numprint{104.8},\numprint{174.8},\numprint{3256591},\numprint{211.7},\numprint{386.6}
uk-2007-05,\numprint{3735056},*,\numprint{71.0},\numprint{298.5},\numprint{369.5},\numprint{3629214},\numprint{510.0},\numprint{879.5}
webbase-2001,\numprint{869443},\numprint{13.0},\numprint{18.3},\numprint{23.6},\numprint{54.9},\numprint{821131},\numprint{14.8},\numprint{69.7}
asia.osm,\numprint{34851},\numprint{0.8},\numprint{0.3},\numprint{0.1},\numprint{1.2},\numprint{34823},\numprint{0.1},\numprint{1.3}
road\_usa,\numprint{246939},\numprint{2.5},\numprint{1.2},\numprint{0.4},\numprint{4.0},\numprint{246939},\numprint{0.3},\numprint{4.3}
europe.osm,\numprint{14152},\numprint{4.1},\numprint{0.7},\numprint{0.2},\numprint{5.0},\numprint{14096},\numprint{0.1},\numprint{5.2}
rgg26,\numprint{49847428},\numprint{1.0},\numprint{44.2},\numprint{103.7},\numprint{148.9},\numprint{49838810},\numprint{780.7},\numprint{929.6}
rhg,\numprint{16},*,\numprint{44.4},\numprint{20.7},\numprint{65.1},\numprint{0},\numprint{28.0},\numprint{93.0}
del24,\numprint{12901142},\numprint{0.2},\numprint{31.7},\numprint{19.0},\numprint{50.9},\numprint{12877029},\numprint{135.0},\numprint{185.8}
del26,\numprint{51668286},\numprint{0.7},\numprint{86.9},\numprint{88.5},\numprint{176.1},\numprint{51624361},\numprint{611.0},\numprint{787.0}
\end{filecontents*}
\csvreader[head to column names, late after line=\\]{include/ourAlgorithmParallel.csv}{}
    {\graph & \ourSize & \LinearTimeTime & \partitioningTime & \quasiKernelTime & \quasiKernelTotal & \kernelSize & \kernelTime & \kerneltotal}
    \end{tabular}

    \captionof{table}{Kernel sizes and kernelization times for \parfastker{} to reach a
      \quasikernel{} and to reach a full kernel by running our algorithm
      sequentially and without
      stopping the reduction application on the \quasikernel.}
\end{figure*}
\begin{figure*}[!h]
  \centering
\footnotesize
\begin{tabular}{l|rr|rrr|rr}%
  & \multicolumn{2}{c|}{\vcsolver{}} & \multicolumn{3}{c|}{\parfastker} & \multicolumn{2}{c}{Same size comparison}\\
 graph & $|\mathcal{K}|$ & time [s] & $|\mathcal{K}|$ &  time [s] & speedup & time [s] & speedup \\
  \hline
    \begin{filecontents*}{include/parallelQuasiKernelComparisonAkiba.csv}
graph,AkibaSize,AkibaTime,ourSize,ourTimeQuasiKernelTotal,totalSpeedup,sameSizeTime,sameSizeSpeedup
uk-2002,\numprint{241517},\numprint{336.9},\numprint{266328},\numprint{11.8},\numprint{28.6},\numprint{199.5},\numprint{16.9}
arabic-2005,\numprint{574878},\numprint{1033.2},\numprint{628850},\numprint{25.7},\numprint{40.3},\numprint{509.5},\numprint{19.9}
gsh-2015-tpd,\numprint{417031},\numprint{372.3},\numprint{486328},\numprint{31.7},\numprint{11.7},\numprint{139.5},\numprint{4.4}
uk-2005,\numprint{835480},\numprint{541.4},\numprint{901896},\numprint{52.8},\numprint{10.2},\numprint{137.1},\numprint{2.6}
it-2004,\numprint{1602560},\numprint{6749.0},\numprint{1697934},\numprint{151.1},\numprint{44.7},\numprint{4108.9},\numprint{27.2}
sk-2005,\numprint{3200806},\numprint{10010.5},\numprint{3504786},\numprint{174.8},\numprint{57.3},\numprint{2822.2},\numprint{16.1}
uk-2007-05,\numprint{3514783},\numprint{18829.4},\numprint{3735056},\numprint{369.5},\numprint{51.0},\numprint{11828.5},\numprint{32.0}
webbase-2001,\numprint{736842},\numprint{4207.8},\numprint{869443},\numprint{54.9},\numprint{76.7},\numprint{2626.6},\numprint{47.9}
asia.osm,\numprint{15201},\numprint{204.7},\numprint{34851},\numprint{1.2},\numprint{172.3},\numprint{159.5},\numprint{134.3}
road\_usa,\numprint{169808},\numprint{310.0},\numprint{246939},\numprint{4.0},\numprint{77.3},\numprint{91.6},\numprint{22.8}
europe.osm,\numprint{8366},\numprint{302.4},\numprint{14152},\numprint{5.0},\numprint{60.0},\numprint{214.5},\numprint{42.6}
rgg26,\numprint{49590973},\numprint{9887.7},\numprint{49847428},\numprint{148.9},\numprint{66.4},\numprint{1637.8},\numprint{11.0}
rhg,\numprint{0},\numprint{124.0},\numprint{16},\numprint{65.1},\numprint{1.9},\numprint{113.4},\numprint{1.7}
del24,\numprint{12417301},\numprint{4789.5},\numprint{12901142},\numprint{50.9},\numprint{94.2},\numprint{1002.4},\numprint{19.7}
del26,\numprint{49864448},\numprint{20728.7},\numprint{51668286},\numprint{176.1},\numprint{117.7},\numprint{4713.6},\numprint{26.8}
\end{filecontents*}
\csvreader[head to column names, late after line=\\]{include/parallelQuasiKernelComparisonAkiba.csv}{}
    {\graph & \AkibaSize & \AkibaTime & \ourSize & \ourTimeQuasiKernelTotal & \totalSpeedup & \sameSizeTime & \sameSizeSpeedup }
    \end{tabular}

    \captionof{table}{Comparison between \vcsolver{} and \parfastker{}. "Same size comparison" compares the time that the
      algorithm with the smaller kernel size takes to reach the final size of
      the algorithm with the larger kernel.}
\end{figure*}
\vfill
\pagebreak
\end{document}

%% file: makros.tex
\usepackage{amsfonts}

\newcommand{\set}[1]{\left\{ #1\right\}}

\newcommand{\realrange}[2]{\left[#1, #2\right]}

\newcommand{\unitrange}[2]{\realrange{0}{1}}

\newcommand{\Oh}[1]{\mathcal{O}\!\left( #1\right)}

\newcommand{\llabel}[1]{\label{\labelprefix:#1}}
\newcommand{\labelprefix}{} 

\newcommand{\discussionsize}{\small}

\newcommand{\frage}[1]{}

\newenvironment{code}{\noindent
\begin{tabbing}%
\hspace{2em}\=\hspace{2em}\=\hspace{2em}\=\hspace{2em}\=\hspace{2em}\=%
\hspace{2em}\=\hspace{2em}\=\hspace{2em}\=\hspace{2em}\=\hspace{2em}\=%
\kill}{\end{tabbing}}

\newcommand{\labelcommand}{}
\newcommand{\captiontext}{}
\newsavebox{\codeparam}
\newcounter{lineNumber}
\newenvironment{disscodepos}[3]{%
\renewcommand{\labelcommand}{#2}%
\renewcommand{\captiontext}{#3}%
\sbox{\codeparam}{\parbox{\textwidth}{#3}}%
\begin{figure}[#1]\begin{center}\begin{code}\setcounter{lineNumber}{1}}{%
\end{code}\end{center}\caption{\llabel{\labelcommand}\captiontext}\end{figure}}

{\end{disscodepos}}

\newdimen\endofsize\endofsize=0.5em
\def\endofbeweis{~\quad\hglue\hsize minus\hsize
                 \hbox{\vrule height \endofsize width
\endofsize}\par}
